\documentclass[preprint,proceedings]{rmaa}

\usepackage{paralist}

\usepackage{psfrag,color}

\SetYear{2005}
\SetConfTitle{Triggering Relativistic Jets}

\title{The Structure and Dynamics of GRB Jets}

\author{Jonathan Granot\altaffilmark{1}}

\altaffiltext{1}{KIPAC, Stanford University, CA, USA.}

\shortauthor{Granot}
\shorttitle{Jets in Gamma-Ray Bursts}

\fulladdresses{
\item Jonathan Granot: Kavli Institute for Particle Astrophysics and
Cosmology, Stanford University, P.O. Box 20450, MS 29, Stanford, CA
94309, USA (\email{granot@slac.stanford.edu}).
}

\listofauthors{J. Granot}
\indexauthor{Granot, J.}

\abstract{There are several lines of evidence which suggest that the
relativistic outflows in gamma-ray bursts (GRBs) are collimated into
narrow jets. The jet structure has important implications for the true
energy release and the event rate of GRBs, and can constrain the
mechanism responsible for the acceleration and collimation of the jet.
Nevertheless, the jet structure and its dynamics as it sweeps up the
external medium and decelerates, are not well understood. In this
review I discuss our current understanding of GRB jets, stressing
their structure and dynamics.}

\addkeyword{gamma rays: bursts}
\addkeyword{hydrodynamics}
\addkeyword{ISM: Jets and outflows}
\addkeyword{relativity}

\begin{document}
\maketitle

\section{Introduction}
\label{sec:intro}

Gamma-ray bursts (GRBs) are produced by a highly relativistic outflow
from a compact source \citep[for a comprehensive recent review
see][]{Piran05}. Early GRB models featured a spherical outflow, mainly
for simplicity. However, other astrophysical sources of relativistic
outflows such as active galactic nuclei and micro-quasars are in the
form of narrow bipolar jets. One might argue \citep[e.g.,][]{Rhoads97}
that, in analogy to other such sources, GRBs might also be collimated
into narrow jets.

The initial Lorentz factor during the prompt gamma-ray emission is
very high, $\Gamma_0 \gtrsim 100$, and therefore we observe emission
mainly from very small angles, $\theta \lesssim \Gamma_0^{-1} \lesssim
10^{-2}\;$rad, relative to our line of sight. This is a result of
relativistic beaming (i.e. aberration of light), an effect of special
relativity, which causes an emission that is roughly isotropic in the
rest frame of the emitting fluid (as is generally expected under most
circumstances) to be concentrated mostly within an angle of
$\Gamma^{-1}$ around its direction of motion in the lab frame, where
$\Gamma \gg 1$ is the Lorentz factor of the emitting fluid in the lab
frame. For this reason, the prompt gamma-ray emission probes a region
of solid angle $\sim \pi\Gamma_0^{-2}$, or a fraction $\sim
\Gamma_0^{-2}/4 \sim 10^{-7} - 10^{-4.5}$ of the total solid angle,
and cannot tell us whether the outflow occupies a larger solid angle.

Therefore, more direct evidence in favor of jets in GRBs had to await
the discovery of afterglow emission in the X-ray \citep{Costa97},
optical \citep{vanParadijs97}, and radio \citep{Frail97}, that lasts
for days, weeks, and months, respectively, after the GRB.  The
afterglow is believed to be synchrotron emission from the shocked
external medium. As the relativistic outflow expands outwards it
sweeps up the surrounding medium and drives a strong relativistic
shock into it, called the forward shock, while the ejecta are
decelerated by a reverse shock. Eventually, most of the energy is
transfered to the shocked external medium behind the forward shock,
and the flow approaches a spherical self-similar evolution
\citep{BM76}, gradually decelerating as it sweeps up the external
medium.

This is valid not only for an initially spherical outflow, but also
for the interior of a jet with an angular size larger than
$\Gamma_0^{-1}$ \citep[as appears to be the case for GRB jets,
e.g.][]{PK02}, as long as $\Gamma^{-1}$ remains smaller than the
angular size of the jet and the interior of the jet is out of causal
contact with its edges (i.e. before the jet break time). Therefore,
before the jet break time the isotropic equivalent energy of the jet,
$E_{\rm iso}$, is relevant (both for its dynamics and for the
resulting emission), while at very late times as the jet becomes
sub-relativistic and approaches spherical symmetry its true energy,
$E$, is relevant. In the intermediate regime things are more
complicated, and are discussed in this review.

The forward shock is responsible for the long lived afterglow
emission, while the reverse shock produces a shorter lived emission,
that peaks in the optical or NIR on a time-scale of tens of seconds,
when the reverse shock crosses the shell of ejecta \citep[the
``optical flash'', e.g.][]{Akerlof99,SP99a,SP99b,MR99}. The shocked
outflow gradually cools adiabatically and the peak of its emission
shifts to lower frequencies, until after about a day it peaks in the
radio \citep[the ``radio
flare''][]{Kulkarni99b,Frail00,Berger03a}. During the afterglow, the
Lorentz factor $\Gamma$ of the emitting shocked external medium
decreases with time as it accumulates more mass, causing the visible
region of $\theta \lesssim \Gamma^{-1}$ around the line of sight to
increase with time. This enables us to probe the structure of the
outflow over increasingly larger angular scales.

Different lines of evidence suggest that the relativistic outflows in
GRBs are collimated into narrow jets. A compelling, although somewhat
indirect, argument comes from the very high values for the energy
output in gamma rays assuming isotropic emission, $E_{\rm\gamma,iso}$,
that are inferred for GRBs with known redshifts, $z$, which approach
and in one case (GRB~991023) even exceed $M_\odot c^2$. Such extreme
energies in an ultra-relativistic outflow are hard to produce in
models involving stellar mass progenitors. If the outflow is
collimated into a narrow jet (or bipolar jets) that occupies a small
fraction, $f_b\ll 1$, of the total solid angle, then the strong
relativistic beaming due to the very high initial Lorentz factor
($\Gamma_0\gtrsim 100$) causes the emitted gamma rays to be similarly
collimated. This reduces the true energy output in gamma rays by a
factor of $f_b^{-1}$ to $E_\gamma = f_b E_{\rm\gamma,iso}$, thus
significantly reducing the energy requirements. 

Estimates of the energy in the afterglow shock from late time radio
observations when the flow is only mildly relativistic and starts to
approach spherical symmetry \citep[often called ``radio
calorimetry'';][]{FWK00,BKF04,Frail05} typically yield $E_k \sim
10^{51.5}\;$erg which lends some support for the true energy being
significantly smaller than $E_{\rm\gamma,iso}$. One should keep in
mind, however, that these are only approximate lower limits on the
true afterglow energy, and the latter can in principle be much higher
\citep[see, e.g.,][]{EW05}.

Furthermore, there is good (spectroscopic) evidence that at least some
GRBs of the long-soft class occur together (to within a few days) with
a core collapse supernova of Type Ic \citep{Stanek03,Hjorth03}. In
such cases the {\it average} Lorentz factor must be
$\langle\Gamma\rangle \lesssim 2$ for a spherical explosion, since the
accreted mass does not significantly exceed the ejected mass, and only
a fraction of the rest energy of the former can provide the kinetic
energy for the latter. Therefore, only a small fraction of the ejected
mass can reach $\Gamma \gtrsim 100$ which is required in order to
power the GRB, and hydrodynamic analysis \citep{TMM01,PV02} shows that
it would carry a small fraction of the total energy which is
insufficient to account for the high end of the observed values of
$E_{\rm\gamma,iso}$. For a jet the ejected mass can be much smaller
than the accreted mass so that $\langle\Gamma\rangle \gg 1$ is
possible, in addition to the smaller $E_\gamma$ that is implied by the
same observed $E_{\rm\gamma,iso}$.

A more direct line of evidence in favor of a narrowly collimated
outflow comes from achromatic breaks seen in the afterglow light
curves of many GRBs
\citep{Fruchter99,Kulkarni99a,Harrison99,Stanek99,Stanek01,
Berger00,Halpern00,Price01,Sagar01,Jensen01}. In fact, such a ``jet
break'' in the afterglow light curve was predicted before it was
detected \citep{Rhoads97,Rhoads99,SPH99}. The cause of the jet break
in the light curve is discussed in \S \ref{subsec:break}.

The properties of GRB jets are of fundamental importance since they
pertain to the GRB energy release, event rate, and the progenitor
model through its ability to produce a particular jet structure. In
particular, a good understanding of the jet structure and dynamics are
crucial in order to reliably address these vital issues. This review
focuses on the dynamics of the jet as it sweeps up the external medium
and decelerates (\S \ref{sec:dyn}) and on its angular structure (\S
\ref{sec:str}), stressing the constraints that may be derived from
various observations. The conclusions are discussed in \S
\ref{sec:conc}.

\section{The Jet Dynamics}
\label{sec:dyn}

This section begins by presenting three different approaches to the
calculation of the jet dynamics, in order of increasing complexity:
semi-analytic models (\S~\ref{subsec:semi-analytic}), simplifying the
dynamical equations by integrating over the radial profile of the jet
(\S~\ref{subsec:2D1D}), and full hydrodynamic simulations
(\S~\ref{subsec:sim}). The main results of the different approaches
are described and compared. Next (\S~\ref{subsec:emission}) there is a
brief description of the typical assumptions that are made in order to
calculate the afterglow emission. The afterglow image is discussed in
\S~\ref{subsec:image} along with potential methods for resolving it
and constraining its angular size, as well as how its morphology and
the evolution of its size may help us learn about the jet dynamics and
the external density profile. Finally, the cause of the jet break in
the afterglow light curve is discussed in \S~\ref{subsec:break}.

\subsection{Simple Semi-Analytic Models}
\label{subsec:semi-analytic}

The first approach that had been adopted for calculating the jet
dynamics was using a simple semi-analytic model
\citep{Rhoads97,Rhoads99}. Many different variations on this basic
approach have followed \citep[e.g.,][]{SPH99,PM99,KP00,MSB00,ONP04}. For
simplicity we present here an analysis that largely follows the model
of \citet{Rhoads99}, and which captures the main features of this type
of models.

The basic underlying model assumptions are (i) a uniform jet within a
finite half-opening angle $\theta_j$ with an initial value $\theta_0$
that has sharp edges, (ii) the shock front is part of a sphere at any
given lab frame time and the emitting fluid behind the forward shock
has a negligible width, (iii) the outer edge of the jet is expanding
sideways at a velocity $c_s \sim c$ in the local rest frame of the
jet, (iv) the jet velocity is always in the radial direction and
$\theta_j \ll 1$. Under these assumptions, the jet dynamics are
obtained by solving the 1D ordinary differential equations for the
conservation of energy and particle number.\footnote{For the adiabatic
energy conserving evolution considered here, the equation for momentum
conservation is trivial in spherical geometry, and does not constrain
the dynamics. For a narrow ($\theta_j \ll 1$) highly relativistic
($\Gamma \gg 1$) jet, the equation for the conservation of linear
momentum in the direction of the jet symmetry axis is almost identical
to the energy conservation equation. When the jet becomes
sub-relativistic the conservation of energy and linear momentum force
it to approach spherical symmetry, and once it becomes quasi-spherical
then again the momentum conservation equation becomes irrelevant.}
The lateral expansion velocity in the comoving frame, $c_s$, is
usually identified with the sound speed, in which case $c_s \approx
c/\sqrt{3}$ while the jet is relativistic. However, this does not have
to be the case: it could in principle be either much smaller ($c_s \ll
c$), or as large as the thermal speed \citep[i.e. $c_s \approx c$
while the jet is relativistic;][]{SPH99}.

The lateral size of the jet, $R_\perp$, and its radius, $R$, are
related by $R_\perp \approx \theta_j R$. We have
\begin{equation}\label{dR_perp}
dR_\perp \approx \theta_j dR + c_sdt' \approx 
\left(\theta_j + \frac{c_s}{c\Gamma}\right)dR\ ,
\end{equation}
where $dt' = dt_{\rm lab}/\Gamma \approx dR/c\Gamma$ and 
\begin{equation}\label{dtheta_dR}
\frac{d\theta_j}{dR} \approx 
\frac{1}{R}\left(\frac{dR_\perp}{dR} - \theta_j\right)
\approx \frac{c_s}{cR\,\Gamma(R)}\ .
\end{equation}
Eq. \ref{dtheta_dR} suggests that $\theta_j \sim \theta_0 +
c_s/c\Gamma$, and therefore the jet expands significantly when
$\Gamma$ drops to $\sim c_s/c\theta_0$. This can occur after the edge
of the jet becomes visible (when $\Gamma \sim \theta_0^{-1}$) for $c_s
< c$ \citep{PM99}. Once the jet begins to expand sideways
significantly, then to zeroth order $\theta_j \propto \Gamma^{-1}$ and
therefore energy conservation suggests that $R \sim {\rm const}$,
since $E \sim \Gamma^2\theta_j^2R^3\rho_{\rm ext}(R)c^2$. Here
$\rho_{\rm ext} = AR^{-k}$ is the external density, which is assumed
to be a power law in radius.\footnote{We consider here and throughout
this review only $k < 3$ for which the shock Lorentz factor decreases
with radius for a spherical adiabatic blast wave during the
self-similar stage of its evolution \citep{BM76}.}  As is shown below,
a more careful analysis shows that $\Gamma\theta_j$ slowly decreases
with radius (Eq. \ref{Gamma_theta_j}) while $\theta_j$ grows very
rapidly with radius (Eq. \ref{theta_scaling}).

\begin{figure}[!t]
\includegraphics[width=1.0\columnwidth]{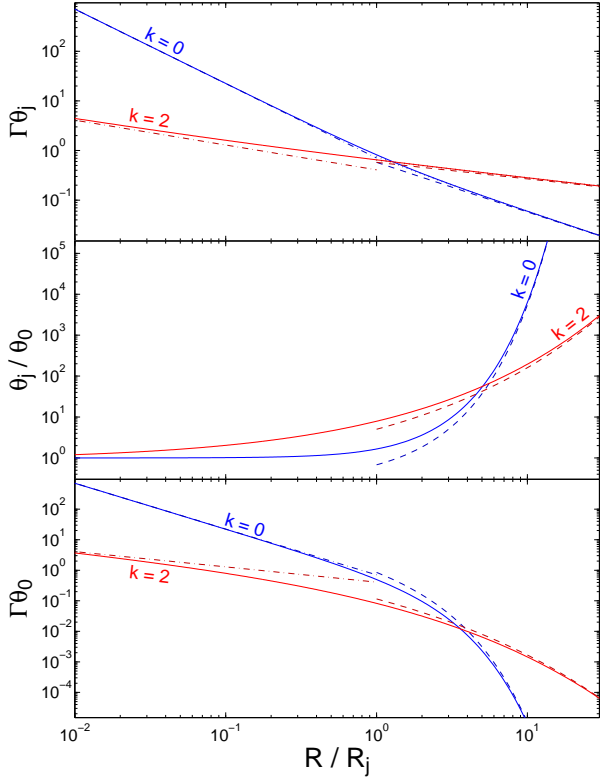}
\caption{The jet dynamics according to the simple semi-analytic model
  that is described in the text ({\it solid lines}), that is obtained
  by numerically solving equations (\ref{dtheta_dRt}) and
  (\ref{dGamma_dRt}) with the initial conditions given by
  $\theta_j(\tilde{R}_0) = \theta_0$ and equation (\ref{Gamma_0}). We
  have used $\beta_s = c_s/c = 3^{-1/2}$, which corresponds the the
  sound speed of a relativistically hot fluid, and show results for a
  uniform external medium ($k = 0$) and for a stellar wind ($k =
  2$). Also shown are the analytic approximations for $R_{\rm dec} < R
  < R_j$ ({\it dashed-dotted lines}) and for $R_j < R < R_{\rm NR}$
  ({\it dashed lines}, according to equations [\ref{Gamma_theta_j}],
  [\ref{theta_scaling}] and [\ref{Gamma_scaling}] with $b = 1/4$). For
  $\Gamma\theta_j$ at $R_j < R < R_{\rm NR}$ we also show (by the {\it
  dotted line}) the higher order approximation given in footnote
  \ref{Gthj}.}
\label{fig:jet_dyn_1D}
\end{figure}

The total swept-up (rest) mass, $M(R)$, is accumulated as
\begin{equation}\label{dM_dR}
\frac{dM}{dR} \approx 2\pi(\theta_jR)^2\rho_{\rm ext}(R) 
=  2\pi AR^{2-k}\theta_j^2(R)\ ,
\end{equation}
where the factor of $2$ is since a double sided jet is assumed. As
long as the jet is relativistic, energy conservation takes the form $E
\approx \Gamma^2 Mc^2$, which implies that $Md(\Gamma^2) =
-\Gamma^2dM$, and
\begin{equation}\label{dGamma_dR}
\frac{d\Gamma}{dR} = -\frac{\Gamma}{2M}\frac{dM}{dR} = -\pi A
R^{2-k}\theta_j^2(R)\frac{\Gamma(R)}{M(R)}\ .
\end{equation}
One can numerically integrate equations (\ref{dtheta_dR}),
(\ref{dM_dR}), and (\ref{dGamma_dR}) thus obtaining $\theta_j(R)$,
$M(R)$, and $\Gamma(R)$. Alternatively, one can use the relation $E
\approx \Gamma^2 Mc^2$ (energy conservation) which reduces the number
of free variable to two, and solve equations (\ref{dtheta_dR}) and
(\ref{dGamma_dR}). Changing variables to a dimensionless radius,
$\tilde{R} \equiv R/R_j$, where
\begin{equation}\label{R_j}
R_j = \left(\frac{E}{\pi A c_s^2}\right)^{1/(3-k)}\ ,
\end{equation}
gives
\begin{eqnarray}\label{dtheta_dRt}
\frac{d\theta_j}{d\tilde{R}} &=&
\frac{\beta_s}{\tilde{R}\,\Gamma(\tilde{R})}\ ,
\\ \label{dGamma_dRt}
\frac{d\Gamma}{d\tilde{R}} &=& 
-\beta_s^{-2}\tilde{R}^{2-k}\theta_j^2(\tilde{R})\Gamma^3(\tilde{R})\ .
\end{eqnarray}
The initial conditions at some small radius $\tilde{R}_0 \ll 1$ (just
after the deceleration radius) are $\theta_j(\tilde{R}_0) = \theta_0$
and
\begin{equation}\label{Gamma_0}
\Gamma(\tilde{R}_0) = 
\sqrt{\frac{3-k}{2}}\,\frac{\beta_s}{\theta_0}\,\tilde{R}_0^{-(3-k)/2}\ .
\end{equation}

Equations (\ref{dtheta_dRt}) and (\ref{dGamma_dRt}) imply,
\begin{equation}\label{dGamma_theta_j_dRt}
\frac{d(\Gamma\theta_j)}{d\tilde{R}} \approx \frac{\beta_s}{\tilde{R}}
-\frac{\tilde{R}^{2-k}}{\beta_s^2}(\Gamma\theta_j)^3 \ .
\end{equation}
If one assumes that the first term becomes dominant at $\tilde{R} > 1$
then this equation implies $\Gamma\theta_j \approx
\beta_s\ln\tilde{R}$, which in turn implies that the second term would
be dominant, rendering the original assumption inconsistent.  The same
applies if the opposite assumption is made, that the second term is
dominant (in this case $\Gamma\theta_j \propto \tilde{R}^{(k-3)/2}$
which implies that the first term would be dominant). This implies
that the two terms must remain comparable, and therefore\footnote{In
order to satisfy equation (\ref{dGamma_theta_j_dRt}) another term with
a smaller power in $\tilde{R}$ is required, $\Gamma\theta_j \approx
\beta_s[\tilde{R}^{(k-3)/3}+\tilde{R}^{2(k-3)/3}(3-k)/9]$, but only
the leading term is shown in equation (\ref{Gamma_theta_j}). This
result is consistent with equation (8) of \citet{KP00} where the
second term in that equation dominates in the relevant
regime.\label{Gthj}}
\begin{equation}\label{Gamma_theta_j}
\Gamma\theta_j \approx \beta_s \tilde{R}^{-(3-k)/3}\ .
\end{equation}
A similar conclusion can be reach by taking the ratio of equations
(\ref{dtheta_dRt}) and (\ref{dGamma_dRt}) which implies that
\begin{equation}
d(\Gamma^{-3}) = \beta_s^{-3}\tilde{R}^{3-k}d(\theta_j^3)\ .
\end{equation}

Substituting equation (\ref{Gamma_theta_j}) into equation
(\ref{dtheta_dRt}) yields
\begin{equation}
\frac{d\theta_j}{d\tilde{R}} \approx
\theta_j\tilde{R}^{-k/3}\ ,
\end{equation}
and
\begin{eqnarray}\label{theta_scaling}
\theta_j &\approx& b\theta_0
\exp\left[\frac{3}{(3-k)}\tilde{R}^{(3-k)/3}\right]\ , 
\\ \label{Gamma_scaling} 
\Gamma &\approx& \frac{\beta_s}{b\theta_0}\tilde{R}^{(k-3)/3}
\exp\left[-\frac{3}{(3-k)}\tilde{R}^{(3-k)/3}\right]\ , 
\quad\quad
\end{eqnarray}
where $b \approx 1/4$ is determined numerically.

\begin{figure}[!t]
\includegraphics[width=1.0\columnwidth]{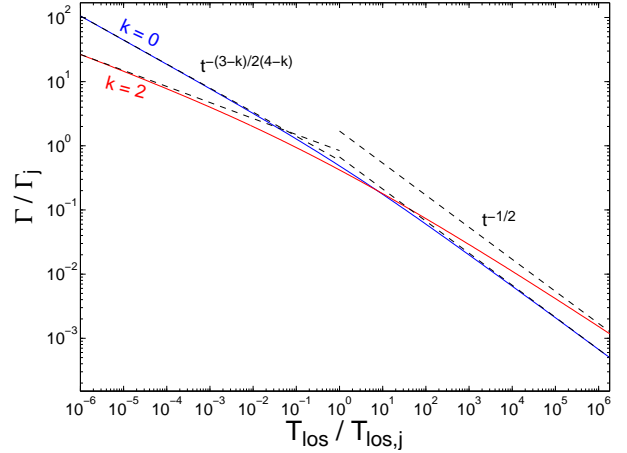}
\caption{The jet Lorentz factor $\Gamma$ as a function of the observed
arrival time of photons emitted along the line of sight $T_{\rm los}
\approx \int dR/2c\Gamma^2$, for the simple semi-analytic model
illustrated in Figure \ref{fig:jet_dyn_1D} ({\it solid lines}). Both
$\Gamma$ and $T_{\rm los}$ are normalized to their values at $R_j$
extrapolated from $R \ll R_j$ ($\Gamma_j$ and $T_{\rm los,j}$,
respectively). Also shown are the asymptotic scalings at $T_{\rm los}
\ll T_j$ and $T_{\rm los} \gg T_j$.}
\label{fig:jet_Gamma_t}
\end{figure}

The results of this simple semi-analytic model are illustrated in
Figures \ref{fig:jet_dyn_1D} and \ref{fig:jet_Gamma_t}. In practice,
the dynamical range between the onset of the exponential lateral
spreading of the jet (at $R_j$) and the non-relativistic transition
(at $R_{\rm NR}$) is quite limited. This fact is ignored in these
figures, and a wide dynamical range is shown in order to better
isolate the characteristics of this intermediate stage ($R_j < R <
R_{\rm NR}$). The dynamical transition at $R \sim R_j$ is much more
gradual for a wind environment ($k = 2$) compared to a uniform density
medium ($k = 0$). This leads to a much smoother and more gradual jet
break in the afterglow light curve \citep{KP00} which would be hard to
detect.

The results derived here are somewhat different than those of
\citet{Rhoads99} who obtained $\Gamma \propto \exp(-\tilde{R})$ and
$\theta_j \propto \tilde{R}^{-1}\exp(\tilde{R})$ at $R_j < R < R_{\rm
NR}$ for a uniform external medium ($k=0$), and they are closer
(though not identical\footnote{The difference arises since there it
was assumed that $M(R) \propto \rho_{\rm ext}(R)R_\perp^2 R$, while
here the differential form is used, $dM \propto \rho_{\rm
ext}(R)R_\perp^2 dR$.}) to those of \citet{Piran00}. This demonstrates
the sensitivity of such semi-analytic models to the exact assumptions
that are made. Nevertheless, despite the differences in their details,
all of these semi-analytic models for the jet dynamics share a similar
main prediction: a very fast lateral expansion (where the jet half
opening angle $\theta_j$ typically grows exponentially with the radius
$R$) after the jet break time. As is discussed in \S \ref{subsec:2D1D}
and \S \ref{subsec:sim}, more detailed numerical calculations of the
jet dynamics, which better capture the relevant physics, contradict
this result and show that the degree of lateral expansion is very
modest as long as the jet is relativistic. Therefore, a simple and
useful approximation for (semi-) analytic calculations would be that
the jet does not expand sideways altogether, retaining its original
opening angle and evolving as if it were part of a spherical flow, as
long as it is relativistic.

\subsection{Intermediate Approach: Integrating over the Radial Profile}
\label{subsec:2D1D}

\begin{figure}[!t]
\vspace{-0.47cm}
\hspace{-0.68cm}
\includegraphics[width=1.15\columnwidth]{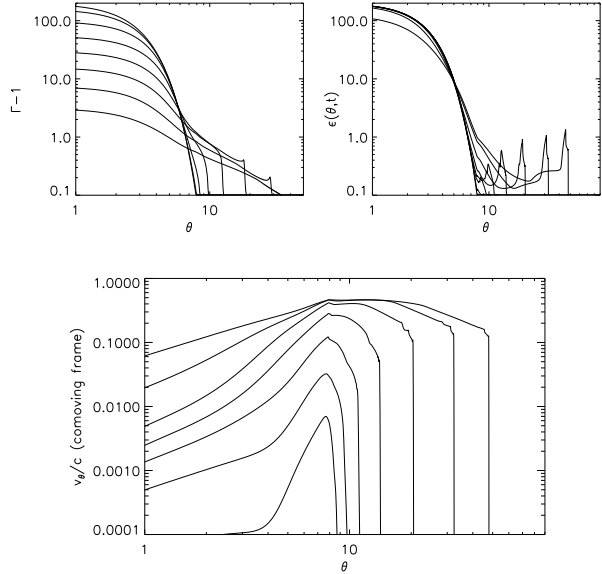}
\\
\vspace{-1.05cm}
\\
\caption{The dynamics of a jet with an initially (at a lab frame time
  $t_0$) Gaussian profile where the energy per solid angle $\epsilon$
  and the Lorentz factor $\Gamma$ are given by
  $\epsilon,\,\Gamma-1\propto\exp(-\theta^2/2\theta_c^2)$, calculated
  using a scheme where the dynamical equations are simplified by
  integrating over the radial profile of the shocked fluid
  \citep{KG03}. The parameters used in this calculation are $\theta_c
  = 0.035\;$rad, $\Gamma(\theta = 0,t_0) = 200$,
  $\epsilon(\theta=0,t_0) = 10^{53}/4\pi\;{\rm erg/sr}$, $n_{\rm ext}
  = 10\;{\rm cm^{-3}}$. The different panels show the evolution of
  $\Gamma-1$, $\epsilon$, and the comoving lateral velocity
  $v'_\theta/c = \Gamma v_\theta/c$. The different curves are for
  different lab frame times: $v'_\theta/c$ increases with time, while
  $\epsilon$ and $\Gamma-1$ decrease with time at the center of the
  jet and increase with time at the sides of the jet behind what
  appears as a shock in the lateral direction which propagates to
  larger angles $\theta$.}
\label{fig:dyn_Gaussian_jet}
\end{figure}

The over-simplified treatment of the jet dynamics in simple
semi-analytic models, and the fact that different such models obtained
different results, put into question the validity of those results and
motivated more careful studies of the jet dynamics. A proper treatment
of this problem requires a full hydrodynamic simulation (in at least
2D) and is discussed in the next subsection. However, since such
simulations are very challenging numerically, an intermediate approach
between simple semi-analytic models and full hydrodynamic simulations
can be useful. This was attempted by \citet{KG03} and is briefly
described here. Under the assumption of axial symmetry, the dynamical
equations are reduced to two spatial dimensions. The dynamical
equations can be reduced to a set of one dimensional (1D) partial
differential equations (PDEs), and thus greatly reducing the
computational difficulty involved in solving these equations, by
integrating over the radial profile of the shocked fluid in the
jet. The motivation for this, apart from making the calculations much
easier, is that most of the shocked fluid is concentrated within a
very thin layer behind the shock, of width $\Delta \sim R/10\Gamma^2$
in the lab frame (i.e. the frame of the unperturbed external
medium). This suggests that integrating over the radial profile might
not introduce a very large error in the calculation of the jet
dynamics.

\begin{figure}[!t]
\vspace{-0.47cm}
\hspace{-0.68cm}
\includegraphics[width=1.15\columnwidth]{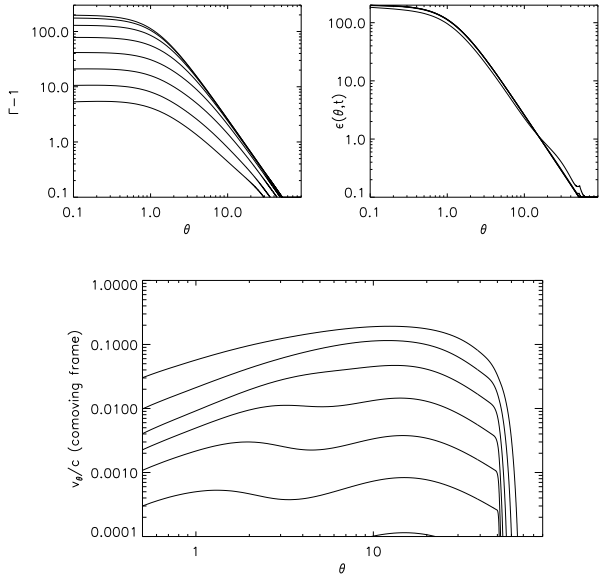}
\\
\vspace{-1.05cm}
\\
\caption{The dynamics of a ``structured'' jet where initially
$\Gamma-1 = 200/(1+\theta^2/\theta_c^2)$ and $\epsilon =
\epsilon_0/(1-\theta^2/\theta_c^2)$ with $\epsilon_0 =
10^{53}/4\pi\;{\rm erg/sr}$, $\theta_c = 0.02\;$rad, and $n_{\rm ext}
= 10\;{\rm cm^{-3}}$. The format is similar to Figure
\ref{fig:dyn_Gaussian_jet}. Again $v'_\theta/c$ increases with time
while $\Gamma-1$ decreases with time. For a structured jet, as opposed
to an initially Gaussian jet, a shock does not develop in the lateral
direction and $\epsilon(\theta)$ remains almost unchanged as long as
the jet is relativistic.}
\label{fig:dyn_structured_jet}
\end{figure}

The results of this method are shown in Figure
\ref{fig:dyn_Gaussian_jet} for a jet with an initial Gaussian profile
in $\Gamma-1$ and in the energy per solid angles $\epsilon =
dE/d\Omega$. Such a jet can be thought of as a smoother and therefore
more realistic version of a uniform jet, since it has a roughly
uniform core and relatively sharp (but still smooth) wings. A shock
appears to develop in the lateral direction, because of the very steep
initial angular profile in the wings of the Gaussian. Nevertheless,
the lateral expansion remains modest as long as the jet is
relativistic. This can be seen both from the small (compared to $c$)
lateral velocity in the comoving frame, and from the fact that
$\epsilon(\theta)$ does not change very much compared to its initial
distribution. The modest degree of lateral spreading is in stark
contrast with the results of semi-analytic models.

Figure \ref{fig:dyn_structured_jet} shows the resulting dynamics for a
``structured'' jet (which is discussed in \S \ref{sec:str}) where
$\Gamma$ and $\epsilon$ are initially power laws with the angle
$\theta$ from the jet symmetry axis, outside of some narrow core.
Again, there is very little lateral expansion (i.e. the comoving
lateral velocity remains $\ll c$ and $\epsilon(\theta)$ hardly
deviates from its initial profile) as long as the jet core is
relativistic.

\subsection{Hydrodynamic Simulations}
\label{subsec:sim}

The most reliable method for calculating the jet dynamics is using
hydrodynamic simulations. This is a formidable numerical task for the
following reasons. First, it requires a hydrodynamic code with special
relativity that is accurate over a large range in the four velocity $u
= \Gamma\beta$, from $u \approx \Gamma \gg 1$ to $u \approx \beta \ll
1$. Second, the shocked fluid in the jet is concentrated in a very
thin layer behind the shock, of width $\Delta \sim R/10\Gamma^2$ in
the lab frame, which is extremely narrow at early times when $\Gamma
\gg 1$, and therefore very hard to resolve properly.

More specifically, from considerations of causality, significant
lateral expansion could in principal occur when $\Gamma$ becomes
comparable to $\theta_0^{-1}$, so that ideally one would want to start
with an initial Lorentz factor $\Gamma_0 \gg \theta_0^{-1}$, and in
practice we need at least $\Gamma_0\theta_0 \gtrsim\;{\rm a\
few}$. Observed jet break times suggest $0.05 \lesssim \theta_0
\lesssim 0.2$ and therefore require $\Gamma_0 \gtrsim 20-100$. 
If $100 N_2$ cells are needed in order to resolve the shell of width
$\Delta \sim R/10\Gamma^2$ then the minimal cell size in the initial
time step (denoted by the subscript `0') needs to be of the order of
$\delta \sim 10^{-6}N_2^{-1}(\Gamma_0/30)^{-2}R_0 \sim
10^{-6}N_2^{-1}(\Gamma_0\theta_0/3)^{-2}(\theta_0/0.1)^2R_0$. The
minimal number of cells required to resolve the initial shell is
$N_{\rm min}\sim \theta_0R_0\Delta_0\delta^{-2} \sim 10^7
N_2^2(\Gamma_0\theta_0/3)^2(\theta_0/0.1)^{-1}$. The total number of
cells in each time step can be $N_{\rm tot} \sim N_{\rm min}$ if the
code uses adaptive mesh refinement (AMR). Otherwise, for a fixed cell
size, $N_{\rm tot}/N_{\rm min} \gtrsim R_0/\Delta_0 \sim
10^4(\Gamma_0\theta_0/3)^2(\theta_0/0.1)^{-2}$.

\begin{figure}[!t]
\hspace{-0.86cm}
\includegraphics[width=1.14\columnwidth]{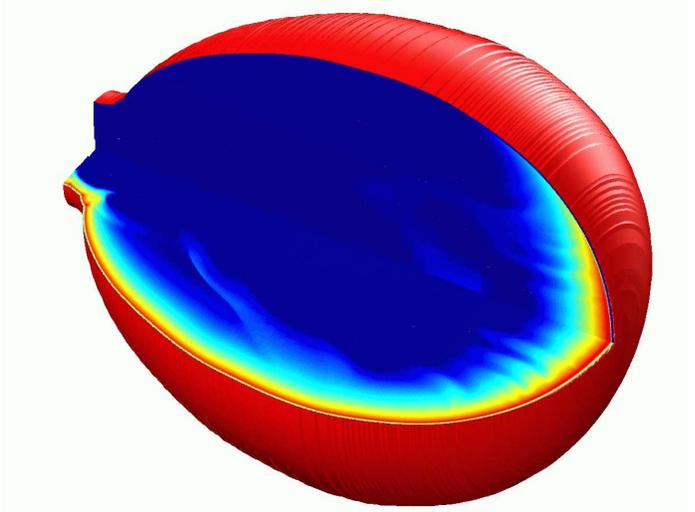}
\caption{A 3D view of a relativistic impulsive jet at the last time
step of the simulation \citep{Granot01}. The outer surface represents
the shock front while the two inner faces show the proper number
density ({\it lower face}) and proper synchrotron emissivity ({\it
upper face}) in a logarithmic color scale.}
\label{fig:jet_dyn_sim}
\end{figure}

\begin{figure}[!t]
\hspace{0.10cm} \includegraphics[width=0.976\columnwidth]{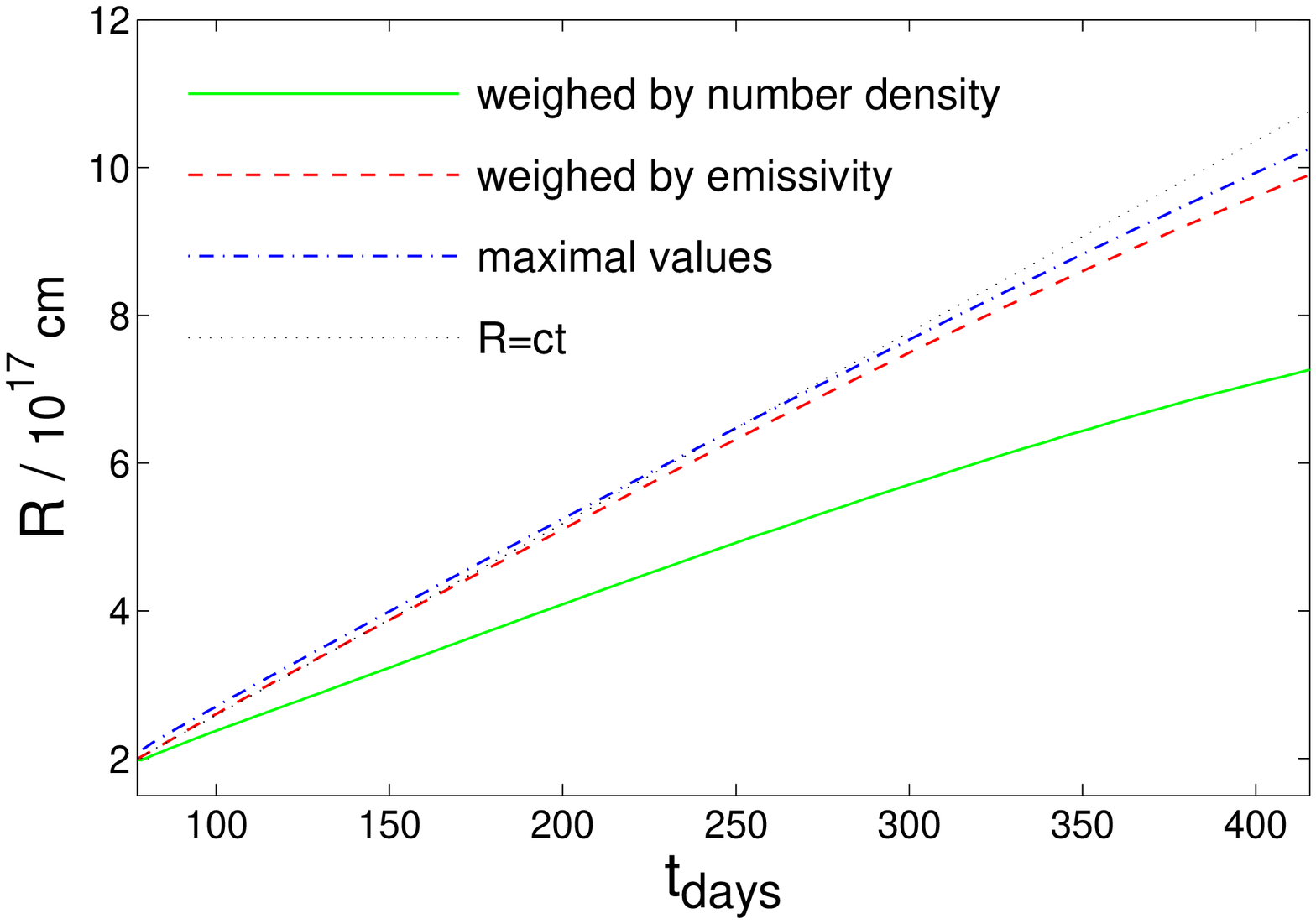} \\
\vspace{0.1cm}
\\
\hspace{0.02cm}
\includegraphics[width=1.005\columnwidth]{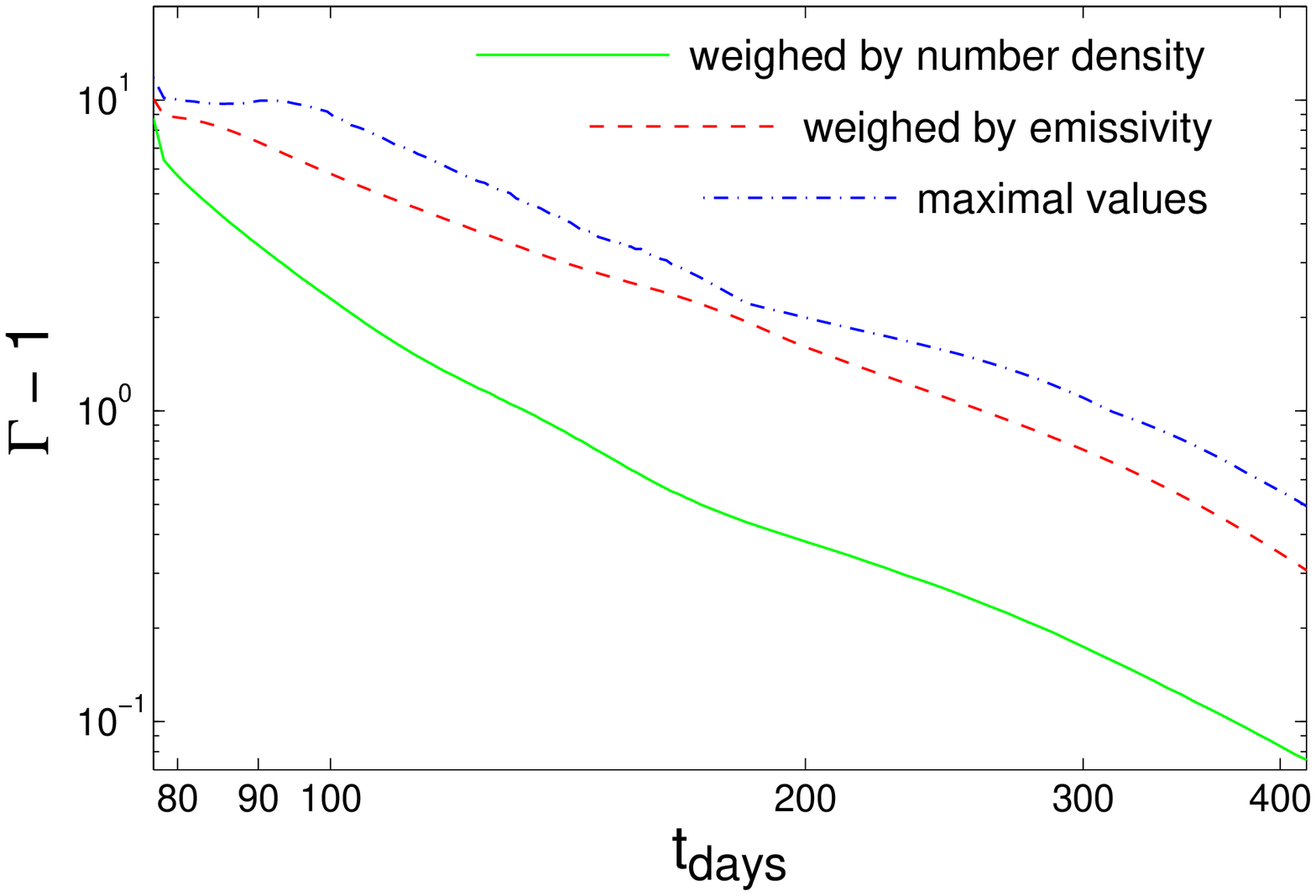}
\\
\vspace{0.1cm}
\\
\hspace{0.2cm}
\includegraphics[width=1.00\columnwidth]{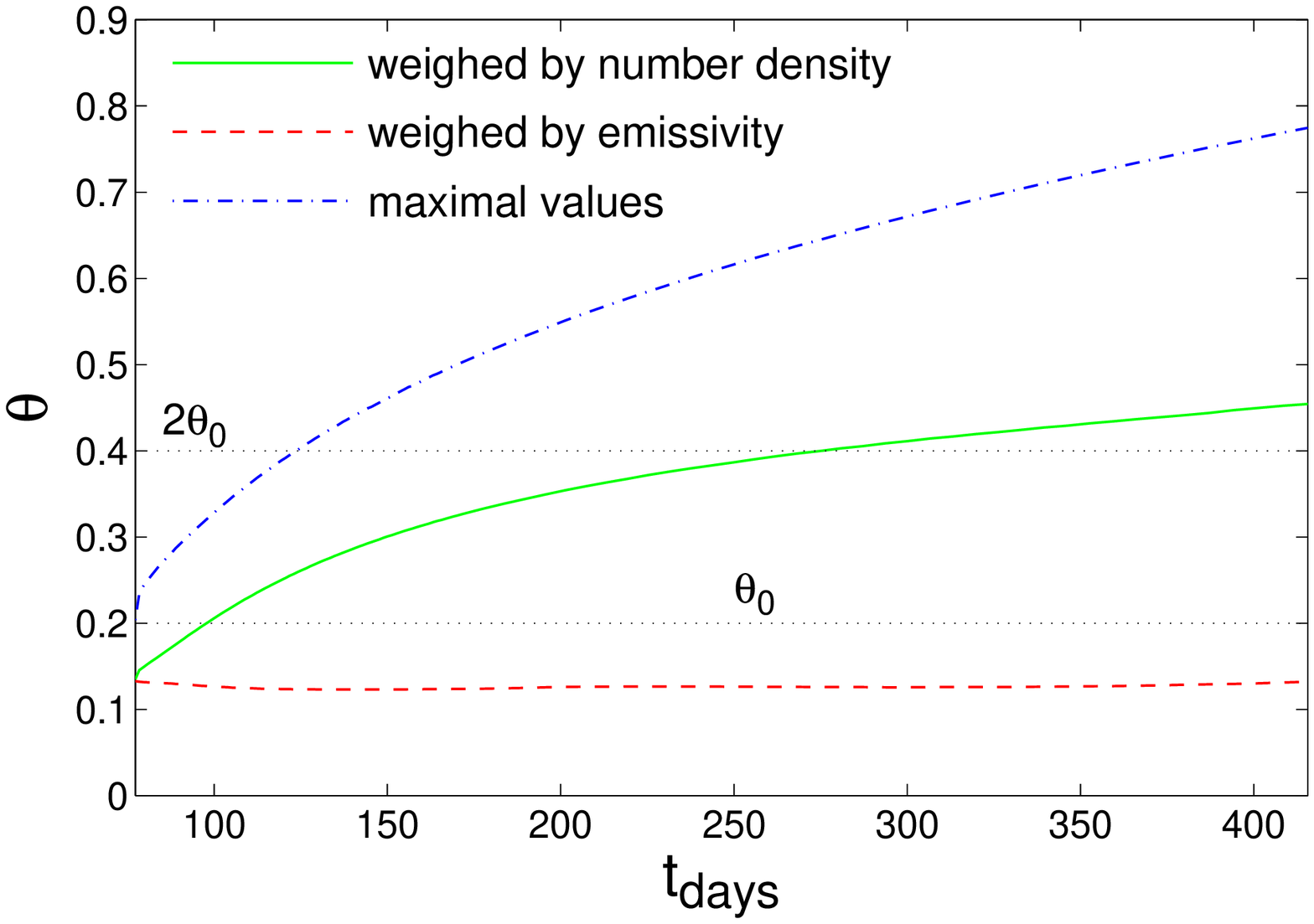}
\\
\vspace{-0.75cm}
\\
\caption{The jet radius ({\it upper panel}), Lorentz factor minus one
  ({\it middle panel}), and angle $\theta$ from the jet symmetry axis
  ({\it lower panel}), as a function of the lab frame time $t_{\rm lab}
  \approx R/c$ (in days), from a hydrodynamic simulation
  \citep{Granot01}.}
\label{fig:jet_dyn_R_g_th}
\end{figure}

Because of the numerical difficulty involved, very few attempts have
been made so far \citep{Granot01,CGV04}. In the following I shall
concentrate on the results of \citet{Granot01}.\footnote{The
calculations of \citet{CGV04} suffer from poor numerical resolution.}
The initial conditions were a cone of half-opening angle $\theta_0 =
0.2\;$rad, taken out of the spherical \citet{BM76} self-similar
solution with an (isotropic equivalent) energy of $E_{\rm k,iso} =
10^{52}\;$erg and a uniform external density of $n_{\rm ext} = 1\;{\rm
cm^{-3}}$. The initial Lorentz factor of the fluid just behind the
shock was $\Gamma_0 \approx 16.8$ corresponding to $\Gamma_0\theta_0
\approx 3.4$. 

The results of the simulation are illustrated in Figures
\ref{fig:jet_dyn_sim} and \ref{fig:jet_dyn_R_g_th}. While the number
density does not change significantly between the front and the sides
of the jet, the emissivity is large only at the front of the jet,
within its initial half-opening angle ($\theta < \theta_0$), and drops
sharply at $\theta > \theta_0$. This causes the emissivity weighted
values of the Lorentz factor and radius to be close to their maximal
values (which are also obtained at the front of the jet).  While the
sides of the jet contribute a small fraction of the total emissivity,
their emission can dominate the observed flux for lines of sight
outside the initial half-opening angle of the jet, as discussed in \S
\ref{subsec:off-axis}. The overall egg-shaped structure of the shock
front is very different from the quasi-spherical structure assumed in
1D semi-analytic models. Moreover, the degree of lateral expansion is
very modest as long as the head of the jet is relativistic, in
contradiction with the very rapid lateral expansion predicted by
semi-analytic models.

\subsection{The Afterglow Emission}
\label{subsec:emission}

The dominant emission mechanism during the afterglow stage is believed
to be synchrotron emission. This is supported by the detection of
linear polarization at the level of $\sim 1\%-3\%$ in several optical
or NIR afterglows (see \S \ref{subsec:lin_pol}), and by the shape of
the broad band spectrum, which consists of several power-law segments
that smoothly join at some typical break frequencies. Synchrotron
self-Compton (SSC; the inverse-Compton scattering of the synchrotron
photons by the same population of relativistic electrons that emits
the synchrotron photons) can sometimes dominate the afterglow flux in
the X-rays \citep{SE01,Harrison01}.

\begin{figure}[!t]
\vspace{0.1cm}
\includegraphics[width=1.00\columnwidth]{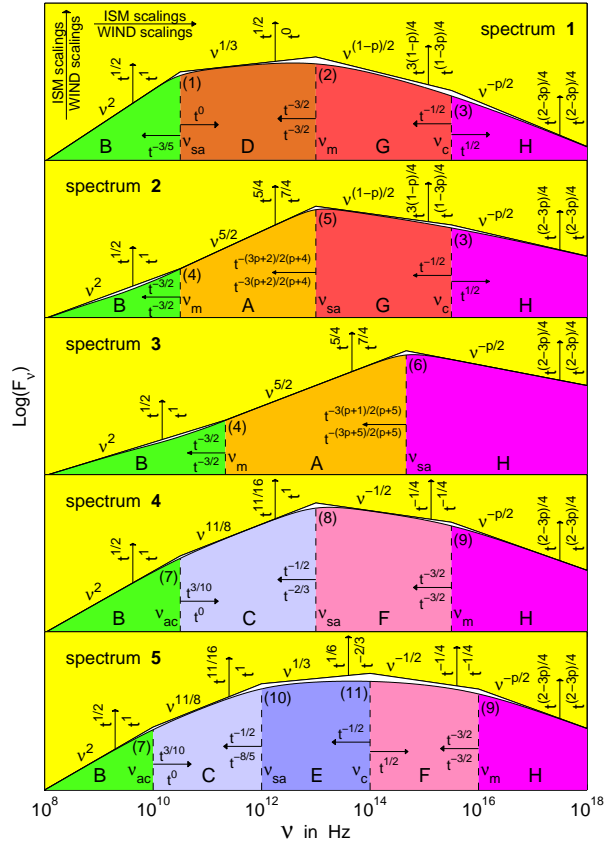}
\caption{The afterglow synchrotron spectrum, calculated for the
\citet{BM76} spherical self-similar solution, under standard
assumptions, using the accurate form of the synchrotron spectral
emissivity and integration over the emission from the whole volume of
shocked material behind the forward (afterglow) shock \citep[for
details see][]{GS02}. The different panels show the five possible
broad band spectra of the afterglow synchrotron emission, each
corresponding to a different ordering of the spectral break
frequencies. Each spectrum consists of several power law segments
(PLSs; each shown with a different color and labeled by a different
letter A--H) that smoothly join at the break frequencies (numbered
1--11). The broken power law spectrum, which consists of the
asymptotic PLSs that abruptly join at the break frequencies (and is
widely used in the literature), is shown for comparison.  Most PLSs
appear in more than one of the five different broad band
spectra. Indicated next to the arrows are the temporal scaling of the
break frequencies and the flux density at the different PLSs, for a
uniform (ISM) and stellar wind (WIND) external density profile.}
\label{fig:spectrum}
\end{figure}

It is usually assumed that the electrons are (practically
instantaneously) shock-accelerated into a power law distribution of
energies, $dN/d\gamma_e \propto \gamma_e^{-p}$ for $\gamma_e >
\gamma_m$, and thereafter cool both adiabatically and due to radiative
losses. Furthermore, it is assumed that practically all of the
electrons take part in this acceleration process and form such a
non-thermal (power-law) distribution, leaving no thermal component
\citep[which is not at all clear or justified; e.g.][]{EW05}. The
relativistic electrons are assumed to hold a fraction $\epsilon_e$ of
the internal energy immediately behind the shock, while the magnetic
field is assumed to hold a fraction $\epsilon_B$ of the internal
energy everywhere in the shocked region. This is a convenient
parameterization of our ignorance regarding the micro-physics of
relativistic collisionless shocks, which are still not sufficiently
well understood from first principals. 

The spectral emissivity in the co-moving frame of the emitting shocked
material is typically approximated as a broken power-law \citep[in
some cases the more accurate functional form of the synchrotron
emission is used, e.g.][]{WG99,GS02}. Most calculations of the light
curve assume emission from an infinitely thin shell, which represents
the shock front \citep[some integrate over the volume of the shocked
fluid taking into account the appropriate radial profile of the flow,
e.g.][see Figure \ref{fig:spectrum}]{GS02}. One also needs to account
for the different arrival times of photons to the observer from
emission at different lab frame times and locations relative to the
line of sight, as well as the relevant Lorentz transformations of the
emission into the observer frame. SSC is included in some (not all)
works, although it can also effect the synchrotron emission through
the enhanced radiative cooling of the electrons.

\subsection{The afterglow Image}
\label{subsec:image}

The apparent surface brightness distribution and size evolution of the
afterglow image on the plane of the sky can potentially provide very
useful information about the structure and dynamics of GRB jets, as
well as about the radial dependence of the external
density. Furthermore, polarimetry (or even spectral polarimetry) of a
resolved afterglow image could provide valuable information on the
magnetic field structure behind collisionless relativistic shocks,
which is not well understood theoretically. However, most GRBs are at
cosmological distances ($z \gtrsim 1$) and the angular size of their
afterglow image is of the order of a micro-arcsecond ($\mu$as) after a
day or so, making it extremely difficult to resolve the image.

\begin{figure}[!t]
\includegraphics[width=1.0\columnwidth]{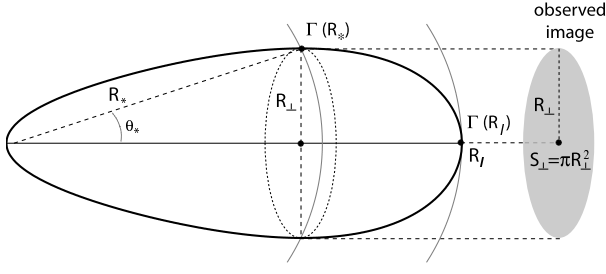}
\caption{Schematic illustration of the equal arrival time surface
  (thick black line), namely, the surface from where the photons
  emitted at the shock front arrive at the same time to the observer
  (on the far right-hand side). The maximal lateral extent of the
  observed image, $R_\perp$, is located at an angle , where the shock
  radius and Lorentz factor are $R_*$ and $\Gamma_* = \Gamma_{\rm
  sh}(R_*)$, respectively. The area of the image on the plane of the
  sky is $S_\perp = \pi R_\perp^2$. The shock Lorentz factor,
  $\Gamma_{\rm sh}$, varies with radius $R$ and angle $\theta$ from
  the line of sight along the equal arrival time surface. The maximal
  radius $R_l$ on the equal arrival time surface is located along the
  line of sight. If, as expected, $\Gamma_{\rm sh}$ decreases with
  $R$, then $\Gamma_l = \Gamma_{\rm sh}(R_l)$ is the minimal shock
  Lorentz factor on the equal arrival time
  surface. \citep[from][]{GR-RL05}.}
\label{fig:im_diag}
\end{figure}

\begin{figure}[!t]
\includegraphics[width=1.00\columnwidth]{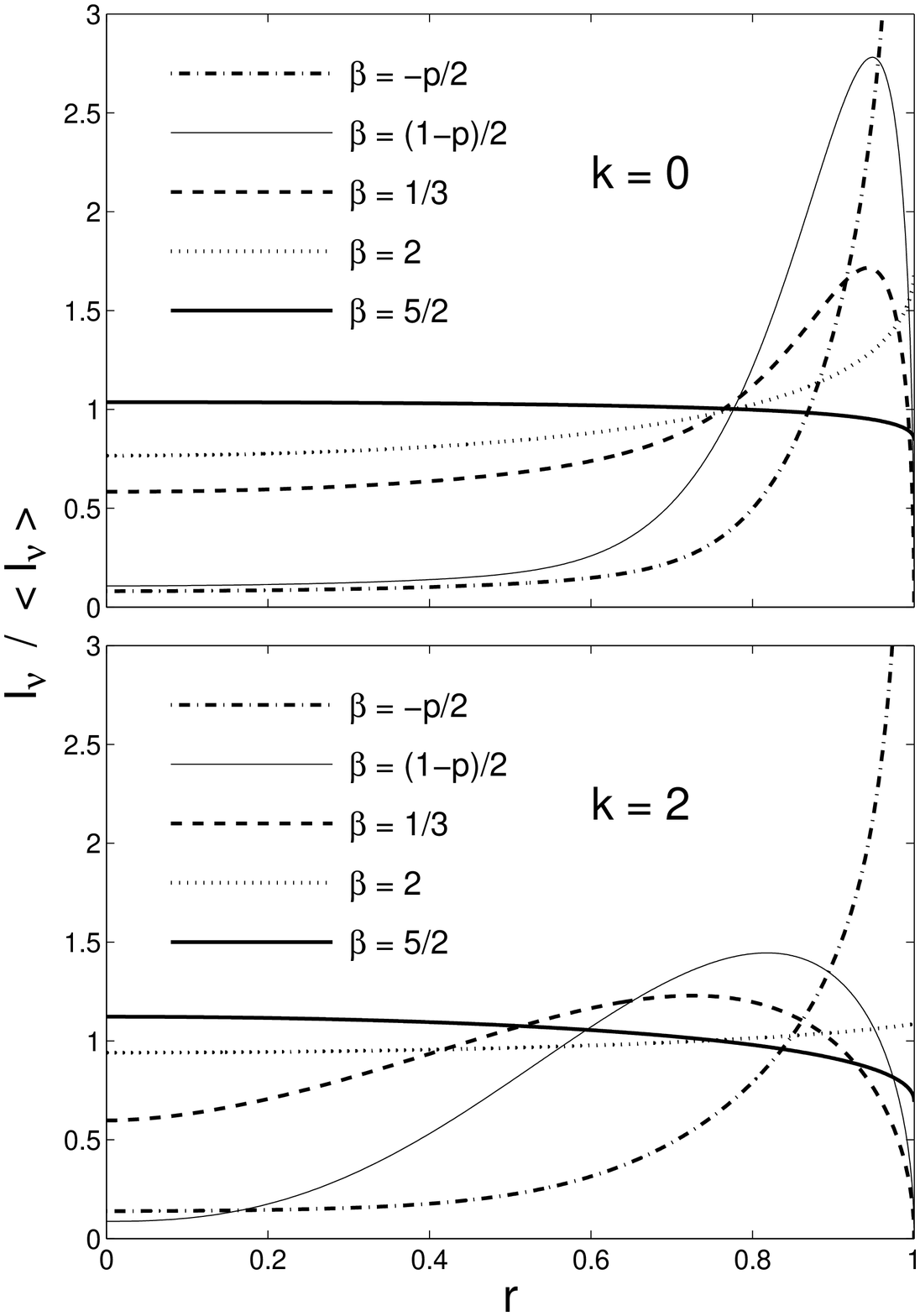}
\caption{The afterglow images for different power law segments of the
  spectrum, for a uniform ($k = 0$) and wind ($k = 2$) external
  density profile \citep[from][calculated for the \citealt{BM76}
  spherical self similar solution, using the formalism of
  \citealt{GS02}]{GL01}.  Shown is the surface brightness, normalized
  by its average value, as a function of the normalized distance from
  the center of the image, $r = R\sin\theta/R_\perp$ (where $r = 0$ at
  the center and $r = 1$ at the outer edge). The image profile changes
  considerably between different power-law segments of the afterglow
  spectrum, $F_\nu \propto \nu^\beta$. There is also a strong
  dependence on the density profile of the external medium, $\rho_{\rm
  ext} \propto R^{-k}$.}
\label{fig:afterglow_images}
\end{figure}

During the self-similar spherical evolution stage (before the jet
break time, for a jet), the afterglow image has circular symmetry
around the line of sight (where the surface brightness depends only on
the distance from the center of the image), and is confined within a
circle on the sky with a radius
\begin{equation}
\frac{R_\perp}{10^{16}\;{\rm cm}} = \left\{\matrix{
  3.91\left(\frac{E_{52}}{n_0}\right)^{1/8}\left(\frac{t_{\rm
      days}}{1+z}\right)^{5/8} & \ \ (k=0) \cr & \cr
  2.39\left(\frac{E_{52}}{A_*}\right)^{1/4}\left(\frac{t_{\rm
      days}}{1+z}\right)^{3/4} & \ \ (k=2)}\right.\ .
\end{equation}
(see Figure~\ref{fig:im_diag}) where $E_{52}$ is the isotropic
equivalent afterglow kinetic energy in units of $10^{52}\;$erg,
$t_{\rm days}$ is the observed time in days, and the external density
is assumed to be a power law with the distance $R$ from the central
source, $\rho_{\rm ext} = n_{\rm ext}m_p = A R^{-k}$, where $n_0 =
n_{\rm ext}/(1\;{\rm cm^{-3}})$ for a uniform external density ($k =
0$), and $A_* = A/(5\times 10^{11}\;{\rm g\;cm^{-1}})$ for a wind-like
external density profile ($k =2$) as might be expected for a massive
star progenitor. This corresponds to an angular radius of
\begin{equation}
\frac{R_\perp}{d_A} = \left\{\matrix{
\frac{1.61\;\mu{\rm
as}}{d_{A,27.7}}\left(\frac{E_{52}}{n_0}\right)^{1/8}
\left(\frac{t_{\rm days}}{1+z}\right)^{5/8} & \ \ (k=0) \cr & \cr
\frac{0.98\;\mu{\rm
as}}{d_{A,27.7}}\left(\frac{E_{52}}{A_*}\right)^{1/4}
\left(\frac{t_{\rm days}}{1+z}\right)^{3/4} & \ \ (k=2)} \right.\ .
\end{equation}
where $d_A(z)$ is the angular distance to the source, and $d_{A,27.7}$
is $d_A$ in units of $10^{27.7}\;{\rm cm} \approx 5\times
10^{27}\;$cm.~\footnote{For a standard cosmology ($\Omega_M = 0.27$,
$\Omega_\Lambda = 0.73$, $h = 0.72$) $d_A(z)$ has a maximum value of
$5.37\times 10^{27}\;$cm ($d_{A,27.7} = 1.07$) for $z = 1.64$.}

\begin{figure}[!t]
\includegraphics[width=0.3270\columnwidth]{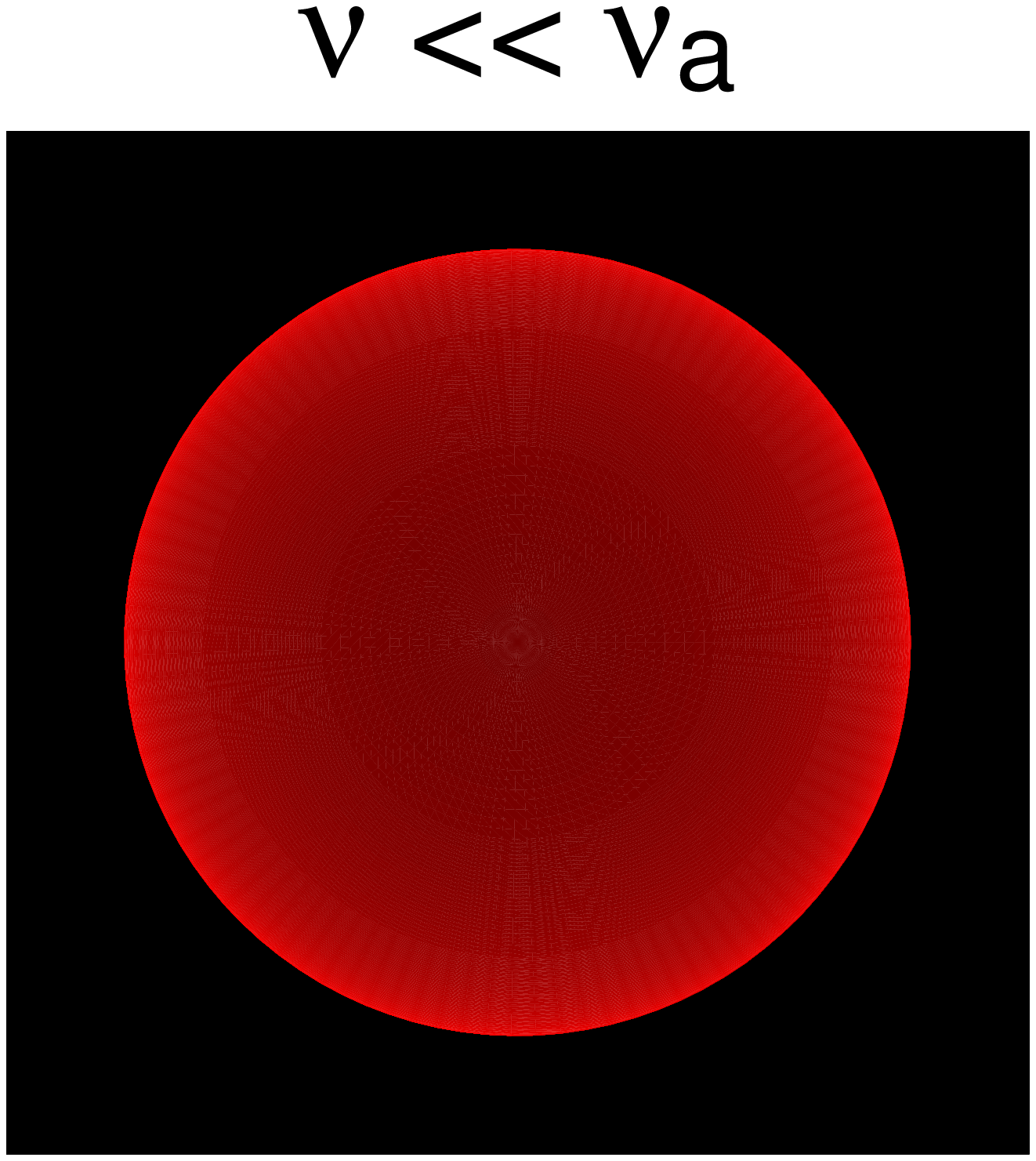}
\includegraphics[width=0.3265\columnwidth]{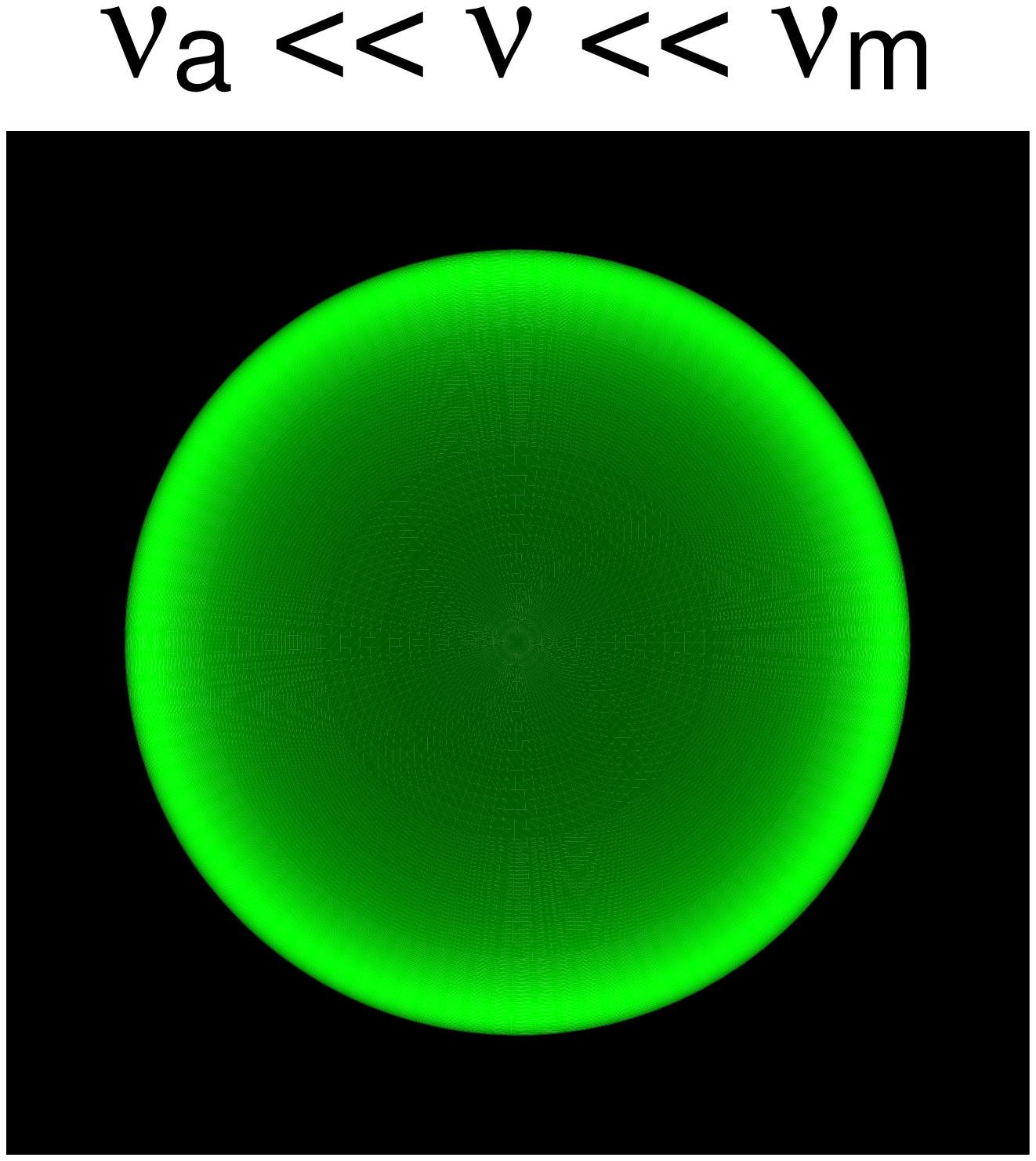}
\includegraphics[width=0.3265\columnwidth]{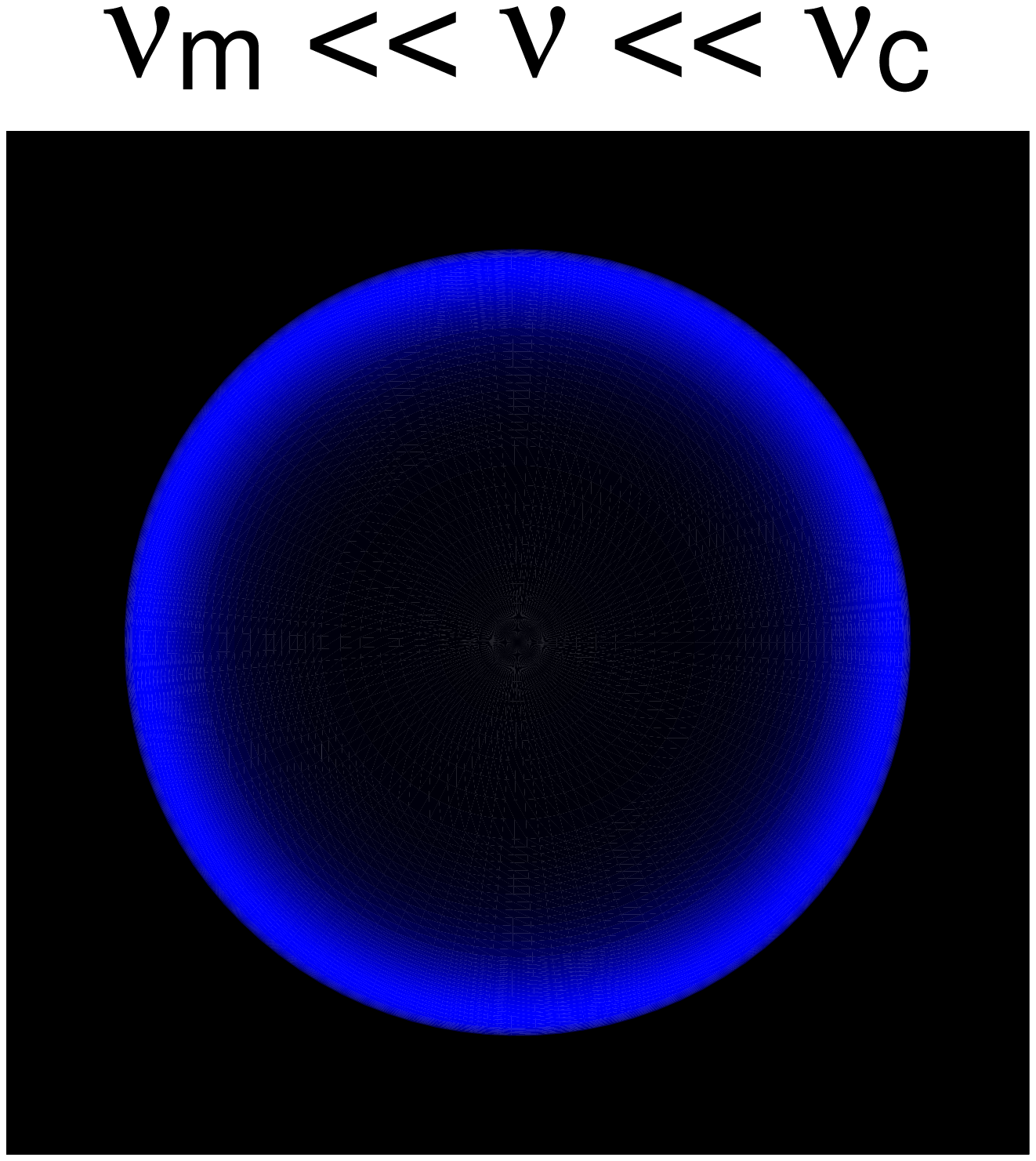}
\\
\vspace{-0.65cm}
\\
\caption{An illustration of the expected afterglow image on the plain
  of the sky, for three different power law segments of the spectrum
  \citep[from][]{GPS99a,GPS99b}, assuming a uniform external density
  and the \citet{BM76} self-similar solution. The image is more limb
  brightened at power law segments that correspond to higher
  frequencies.}
\label{fig:afterglow_images_3}
\end{figure}

More generally, the afterglow image size during the self-similar
spherical stage scales with the observed time as $R_\perp \propto
t^{(5-k)/2(4-k)}$. The image size grows super-luminally with an
apparent expansion velocity of $\Gamma_{\rm sh}(R_*)c$. The expected
afterglow images in this self-similar regime are shown in Figures
\ref{fig:afterglow_images} and \ref{fig:afterglow_images_3}. The
normalized surface brightness profile within the afterglow image is
independent of time due to the self-similar dynamics, and changes only
between the different power law segments of the synchrotron spectrum,
and for different external density profiles. The image becomes
increasingly limb-brightened at higher frequencies, and for smaller
values of $k$. 

Below the self-absorption frequency the specific intensity (surface
brightness) represents the Rayleigh-Jeans portion of a black-body
spectrum with the blue-shifted effective temperature of the electrons
at the corresponding radius along the front side of the equal arrival
time surface of photons to the observer ($R_* \leq R \leq R_l$ in
Figure \ref{fig:im_diag}). Above the cooling break frequency the
emission originates from a very thin layer behind the shock front
where the electrons whose typical synchrotron frequency is close to
the observed frequency have not yet had enough time to significantly
cool due to radiative losses. This results in a divergence of the
surface brightness at the outer edge of the image \citep{Sari98,GL01}.

After the jet break time the afterglow image is no longer symmetric
around the line of sight to the central source for a general viewing
angle (which is not exactly along the jet symmetry axis), and its
details depend on the the hydrodynamic evolution of the jet (so that
in principal it could be used in order to constrain the jet dynamics).
Therefore, a realistic calculation of the afterglow image during the
more complicated post-jet break stage requires the use of
hydrodynamic simulation, and still remains to be done.

The afterglow image may be indirectly resolved through gravitational
lensing by a star in an intervening galaxy (along, or close to, our
line of sight to the source). This is since the angular size of the
Einstein radius (i.e. the region of large magnification around the
lensing star) for a typical star at a cosmological distance is $\sim
1\;\mu$as (hence the name micro-lensing), and therefore comparable to
the afterglow image size after a day or so. Since the afterglow image
size grows very rapidly with time, different parts of the image sample
the regions of large magnification (close to the point of infinite
magnification just behind the lensing star) with time, and therefore
the overall magnification of the afterglow flux as a function of time
probes the surface brightness profile of the afterglow image. This
results in a bump in the afterglow light curve which peaks when the
limb-brightened outer part of the image sweeps past the lensing star,
where the peak of the bump is sharper the more limb-brightened the
afterglow image \citep{GL01}. It has been suggested that an achromatic
bump in the afterglow light curve of GRB~000301C after $\sim 4\;$days
might have been due to micro-lensing \citep{GLS00}. If this
interpretation is true, then the shape of the bump in the afterglow
light curve requires a limb-brightened afterglow image, in agreement
with theoretical expectations \citep{GGL01}.

\begin{figure}[!t]
\includegraphics[width=1.00\columnwidth]{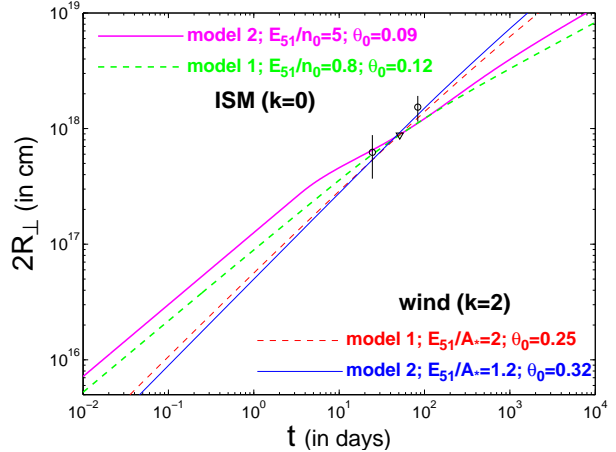}
\caption{Tentative fits to the constraints on the image size (or
  diameter $2R_\perp$) of the radio afterglow of GRB~030329 at
  different epochs, for different external density profiles and
  different assumptions on the lateral spreading of the jet. The
  physical parameters and external density profile for each model are
  indicated (the viewing angle is along the jet symmetry axis). Model
  1 features relativistic sideways expansion in the co-moving rest
  frame of the jet material, while model 2 has no lateral spreading
  \citep[from][]{GR-RL05}.}
\label{fig:im_size}
\end{figure}

The size of the afterglow image at a single epoch can be estimated
from the quenching of diffractive scintillations in the radio
afterglow \citep{Goodman97,Frail97,Taylor97,WKF98}. The flux below the
self-absorption frequency can also be used to constrain the size of
the emitting region \citep[e.g.,][]{KP97,GR-RL05}. A more direct
measurement of the image size, as well as its temporal evolution, may
be obtained through very large base-line interferometry in the radio
(i.e. with the VLBA). This was possible for only for one radio
afterglow so far \citep[GRB~030329;][]{Taylor04,Taylor05}, since it
requires a relatively nearby event ($z \lesssim 0.2$) with a bright
radio afterglow. Nevertheless, it already provides interesting
constraints \citep[][see Figure~\ref{fig:im_size}]{ONP04,GR-RL05}, and
better observations in the future may help pin down the jet structure
and dynamics, as well as the external density profile.

\subsection{What causes the Jet Break?}
\label{subsec:break}

The jet break in the afterglow light curve has been argued to be the
combination of (i) the edge of the jet becoming visible, and (ii) fast
lateral spreading. Both effects are expected to take place around the
same time, when the Lorentz factor, $\Gamma$, of the jet drops below
the inverse of its initial half-opening angle, $\theta_0$. This can be
understood as follows. 

When $\Gamma$ drops below $\theta_0^{-1}$ the edge of the jet becomes
visible, since relativistic beaming limits the region from which a
significant fraction of the emitted radiation reaches the observer to
within an angle of $\sim \Gamma^{-1}$ around the line of sight
($\theta \lesssim \Gamma^{-1}$). Once the edge of the jet becomes
visible, then if there is no significant lateral spreading, only a
small fraction $(\Gamma\theta_j)^2 < 1$ of the visible region is
occupied by the jet, and therefore there would be ``missing''
contributions to the observed flux compared to a spherical flow. This
would cause a steepening in the light curve, i.e. a jet break, where
the temporal decay index asymptotically increases by $\Delta\alpha =
(3-k)/(4-k)$.

\begin{figure}[!t]
\includegraphics[width=1.00\columnwidth]{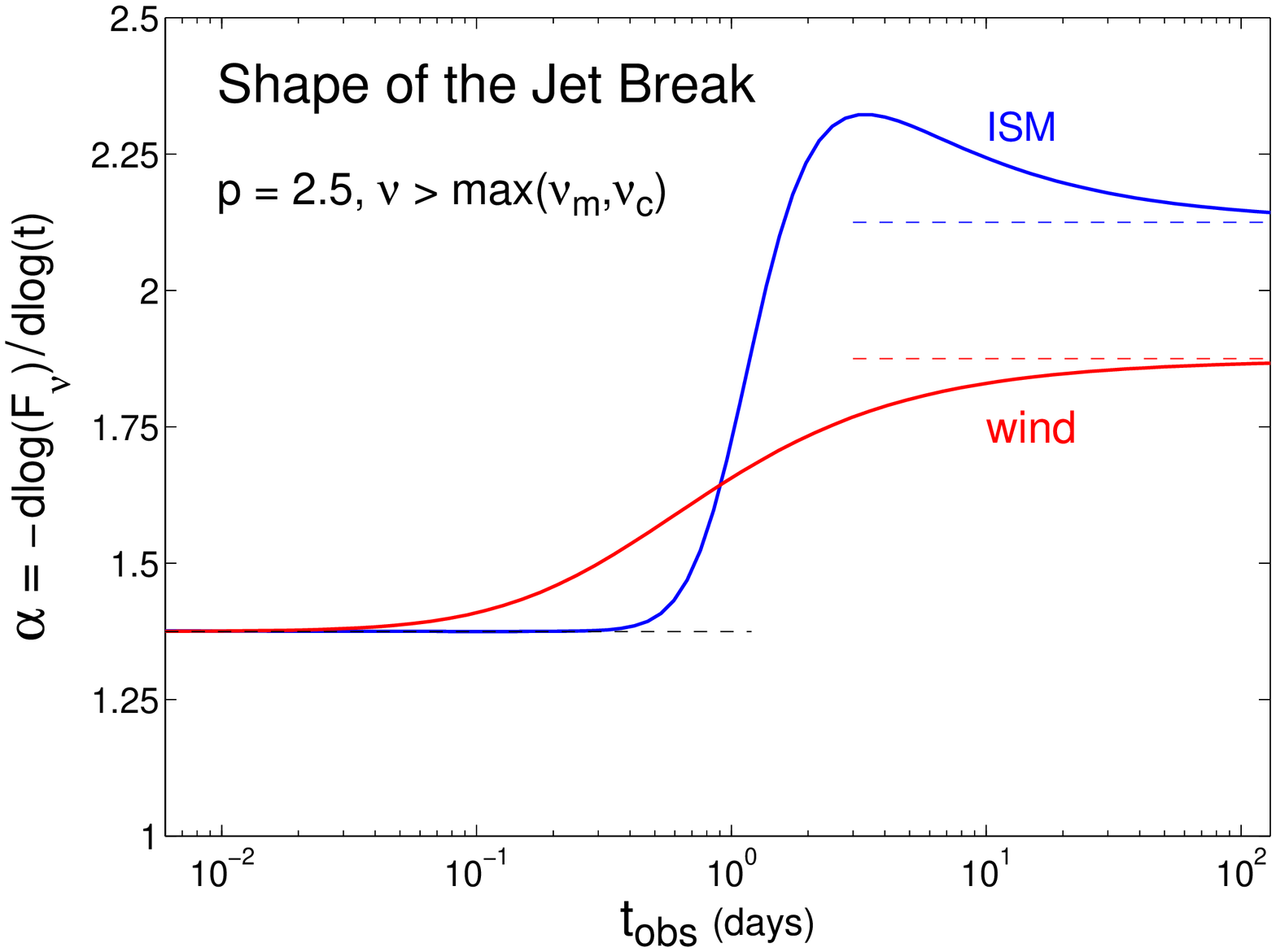}
\\
\vspace{0.1cm}
\\
\includegraphics[width=1.00\columnwidth]{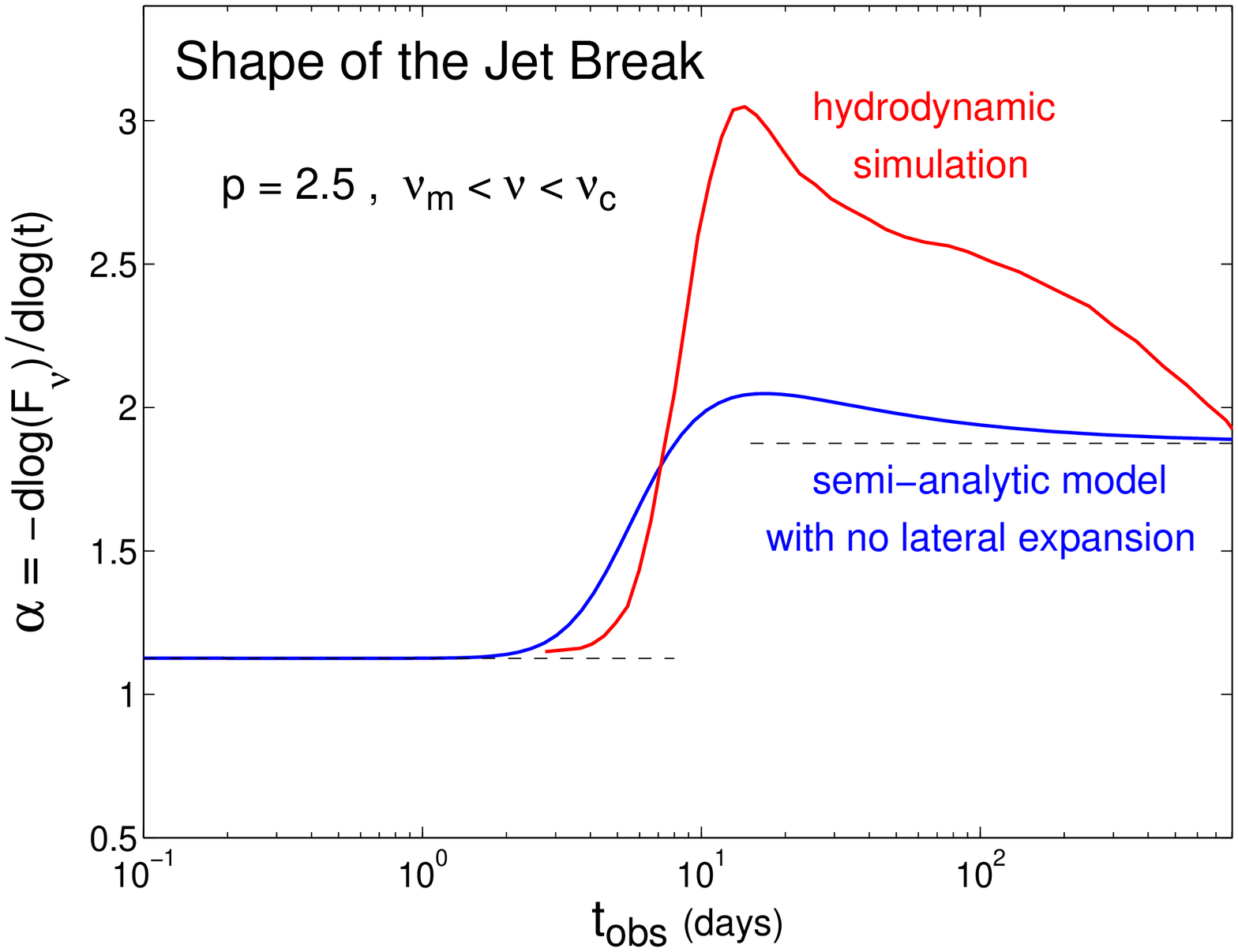} 
\\
\vspace{-0.65cm}
\\
\caption{The temporal decay index $\alpha$ as a function of the
  observed time (in days) across the jet break in the light curve, for
  $p = 2.5$. {\it Upper panel}: results in the spectral range $\nu >
  \max(\nu_m,\nu_c)$ using a semi-analytic model with no lateral
  spreading \citep{Granot05}, for a uniform ($k = 0$, $n_{\rm ext} =
  1\;{\rm cm^{-3}}$) and wind ($k = 2$, $A_* = 1$) external density
  profile, with $\theta_0 = 0.1$ and $E_{\rm k,iso} = 2\times
  10^{53}\;$erg.  {\it Lower panel}: results for the spectral range
  $\nu_m < \nu < \nu_c$, for $\theta_0 = 0.2$ and a uniform density
  ($k = 0$, $n_{\rm ext} = 1\;{\rm cm^{-3}}$, $E_{\rm k,iso} =
  10^{52}\;$erg); compares the result of a semi-analytic model
  \citep{Granot05} to those of a hydrodynamic simulation
  \citep{Granot01}. In both panels the dashed lines show the
  asymptotic values of $\alpha$ before and after the jet break, for a
  uniform jet with no lateral spreading, for which $\Delta\alpha =
  (3-k)/(4-k)$.}
\label{fig:jet_break}
\end{figure}

When $\Gamma$ drops below $\theta_0^{-1}$, the center of the jet comes
into causal contact with its edge, and the jet can in principal start
to expand sideways significantly. It has been argued that at this
stage it would indeed start to expand sideways rapidly, at close to
the speed of light in its own rest frame. In this case, during the
rapid lateral expansion phase the jet opening angle grows as $\theta_j
\sim \Gamma^{-1}$ and exponentially with radius (see \S
\ref{subsec:semi-analytic}). This causes the energy per solid angle,
$\epsilon$, in the jet to drop with observed time, and the Lorentz
factor to decrease faster as a function of the observed time, which
result a steepening in the afterglow light curve compared to a
spherical flow (where $\epsilon$ remains constant and $\Gamma$
decreases more slowly with the observed time). However, in this case a
good part of the visible region remains occupied by the jet (since
$\Gamma\theta_j$ remains $\sim 1$), so that the first cause for the
jet break (the edge of the jet becoming visible, and the ``missing''
contributions from outside the edge of the jet) is no longer
important. Therefore, for fast lateral spreading, the jet break is
caused predominantly since the energy per solid angle $\epsilon$
decreases with time, and the Lorentz factor decreases with observed
time faster than for a spherical flow.

It is important to keep in mind, however, that numerical studies show
that the lateral spreading of the jet is very modest as long as it is
relativistic (see \S \ref{subsec:2D1D} and \S \ref{subsec:sim}). This
implies that lateral spreading cannot play an important role in the
jet break, and the predominant cause of the jet break is the
``missing'' contribution from outside of the jet, once its edge
becomes visible.

A potential problem with this picture is that if the jet half-opening
angle remains roughly constant, $\theta_j \approx \theta_0$, the
asymptotic change in the temporal decay index is only $\Delta \alpha =
3/4$ for a uniform external medium ($k = 0$) or even smaller for a
wind ($\Delta\alpha = 1/2$ for $k = 2$), while the values inferred
from observations are in most cases larger \citep[see Figure 3
of][]{ZKK06}. This apparent discrepancy may be reconciled as follows.
While the {\it asymptotic} steepening is indeed $\Delta \alpha =
(3-k)/(4-k)$ when lateral expansion is negligible, the value of the
temporal decay index $\alpha$ (where $F_\nu \propto t^{-\alpha}$)
initially overshoots its asymptotic value. Since the temporal baseline
that is used in order to measure the post-jet break temporal decay
index $\alpha_2$ is typically no more than a factor of several in time
after the jet break time\footnote{This is usually because the flux
becomes too dim to detect above the host galaxy, or since a supernova
component becomes dominant in the optical, etc.}, $t_{\rm jet}$, the
value of $\alpha$ during this time is {\it larger} than its asymptotic
value $\alpha_2$.  This causes the value of $\Delta\alpha$ that is
inferred from observations to be larger than its asymptotic value.

The overshoot in the value of $\alpha$ just after the jet break time
can nicely be seen in Figure \ref{fig:jet_break}, and is much more
pronounced in the light curves calculated using the jet dynamics from
a hydrodynamic simulation, compared to the results of a simple
semi-analytic model. The cause of this overshoot is that the afterglow
image is limb-brightened (see Figure \ref{fig:afterglow_images}) and
therefore the outer edges of the image which are the brightest are the
first region whose contribution to the observed flux is ``missed'' as
the edge of the jet becomes visible. The overshoot is larger the more
limb-brightened the afterglow image (e.g., for $\nu >
\max(\nu_m,\nu_c)$ in the upper panel of Figure \ref{fig:jet_break}
compared to $\nu_m < \nu < \nu_c$ in the lower panel of Figure
\ref{fig:jet_break}). For a wind density ($k = 2$) the
limb-brightening is smaller compared to a uniform density ($k = 0$),
at the same power law segment of the spectrum (see Figure
\ref{fig:afterglow_images}), and the Lorentz factor $\Gamma$ decreases
more slowly with the observed time. Because of this no overshoot is
seen in the semi-analytic model shown in the upper panel of Figure
\ref{fig:jet_break} for a wind density profile ($k = 2$), and the jet
break is smoother and extends over a larger factor in time. The
asymptotic post-jet break value of the temporal decay index
($\alpha_2$) is approached only when the visible part of the afterglow
image covers the relatively uniform central part, and not its brighter
outer edge.

The jet break in light curves calculated from hydrodynamic simulations
is sharper than in semi-analytic models (where the emission is taken
to be from a 2D surface -- usually a section of a sphere within a
cone). In semi-analytic models the jet break is sharpest with no
lateral expansion, and becomes more gradual the faster the assumed
lateral expansion. For example, in the lower panel of
Figure~\ref{fig:jet_break}, where the viewing angle is along the jet
axis and the external density is uniform, most of the change in the
temporal decay index $\alpha$ occurs over a factor of $\sim 2$ in time
for the numerical simulation, and over a factor of $\sim 3$ in time
for the semi-analytic model (which assumes no lateral expansion; the
jet break would be more gradual with lateral expansion). For both
types of models, the jet break is more gradual and occurs at a
somewhat later time for viewing angles further away from the jet
symmetry axis but still within its initial opening angle, although
this effect is somewhat more pronounced in semi-analytic models
\citep{Granot01,Rossi04}.

\section{The Jet Structure}
\label{sec:str}

Since the initial discovery of GRB afterglows in the X-ray
\citep{Costa97}, optical \citep{vanParadijs97}, and radio
\citep{Frail97}, many afterglows have been detected and the quality of
individual afterglow light curves has improved dramatically
\citep[e.g.,][]{Lipkin04}. Despite all the observational and
theoretical progress, the structure of GRB jets remains largely an
open question. This question is of great importance and interest,
since it is related to issues that are fundamental for our
understanding of GRBs, such as their event rate, total energy, and the
requirements from the compact source that accelerates and collimates
these jets.

In \S \ref{subsec:mod_jet_str} a brief overview is given of the main
jet structures that have been discussed in the literature and the
motivation for them. This is followed by a discussion of the different
methods that have been applied for constraining the jet structure from
observations, which include statistical studies (\S
\ref{subsec:statistics}), as well as the evolution of the linear
polarization of the afterglow emission (\S \ref{subsec:lin_pol}), and
the shape of the afterglow light curves (\S
\ref{subsec:afterglow_LC}).  The afterglow light curves from viewing
angles outside the initial jet aperture are discussed in \S
\ref{subsec:off-axis} along with possible implications for X-ray
flashes and for the jet structure, while \S \ref{subsec:orphans}
briefly mentions the search for orphan afterglows. Some implications
of recent {\it Swift} observations are discussed in \S
\ref{subsec:Swift}.

\subsection{Existing Models for the Jet Structure}
\label{subsec:mod_jet_str}

The leading models for the jet structure are (i) the uniform jet (UJ)
model
\citep{Rhoads97,Rhoads99,PM99,SPH99,KP00,MSB00,Granot01,Granot02},
where the energy per solid angle, $\epsilon$, and the initial Lorentz
factor, $\Gamma_0$, are uniform within some finite half-opening angle,
$\theta_j$, and sharply drop outside of $\theta_j$; and (ii) the
universal structured jet (USJ) model \citep{LPP01,Rossi02,ZM02}, where
$\epsilon$ and $\Gamma_0$ vary smoothly with the angle $\theta$ from
the jet symmetry axis. In the UJ model the different values of the jet
break time, $t_j$ , in the afterglow light curve arise mainly due to
different $\theta_j$ (and to a lesser extent due to different ambient
densities). In the USJ model, all GRB jets are intrinsically
identical, and the different values of $t_j$ arise mainly due to
different viewing angles, $\theta_{\rm obs}$, from the jet
axis.\footnote{In fact, the expression for $t_j$ is similar to that
for a uniform jet with $\epsilon \to \epsilon(\theta=\theta_{\rm
obs})$ and $\theta_j \to \theta_{\rm obs}$} 

The observed correlation, $t_j \propto E_{\rm\gamma,iso}^{-1}$
\citep{Frail01,BFK03}, implies a roughly constant true energy, $E$,
between different GRB jets in the UJ model, and $\epsilon \propto
\theta^{-2}$ outside of some core angle, $\theta_c$, in the USJ model
\citep{Rossi02,ZM02}. This is assuming a constant efficiency,
$\epsilon_\gamma$, for producing the observed prompt gamma-ray (or
X-ray) emission. If the efficiency depends on $\theta$ in the USJ
model, for example, then different power laws of $\epsilon$ with
$\theta$ are possible \citep{GGB05}, such as a core with wings where
$\epsilon \propto \theta^{-3}$, as is obtained in simulations of the
collapsar model \citep{ZWM03,ZWH04}.\footnote{Simulations of jets
launched by an accretion torus - black hole system, where the jet does
not propagate through a stellar envelope, as is expected to arise in
binary merger scenarios which might be relevant to GRBs of the
short-hard class, produce a roughly uniform jet with sharp edges
\citep{AJM05}.}

The jet structure was initially envisioned to be uniform since this is
the simplest jet structure, and arguably also the most
natural. Furthermore, it also predicted a jet break in the afterglow
light curve \citep{Rhoads97,Rhoads99,SPH99}, which was soon thereafter
confirmed observationally \citep{Fruchter99,Kulkarni99a,Stanek99}.
The original motivation for the USJ model was the conceptual
simplicity of a universal intrinsic structure for all GRB jets, where
the observed differences (namely in the jet break times) arise due to
different viewing angles (instead of being attributed to an intrinsic
difference - in the jet half-opening angle - as in the UJ model). Its
exact structure was motivated by the requirement to reproduce the
observed afterglow light curves and correlations with the prompt GRB
emission. It had later been suggested on theoretical grounds that a
jet structure with a narrow core and wings where $\epsilon \propto
\theta^{-2}$ is expected in highly magnetized Poynting flux dominated
jets \citep{LB02,LB03} as well as in low magnetization hydrodynamic
jets in the context of the collapsar model (\citealt{LB05}; see,
however, \citealt{MLB06}).

\begin{figure}[!t]
\includegraphics[width=1.00\columnwidth]{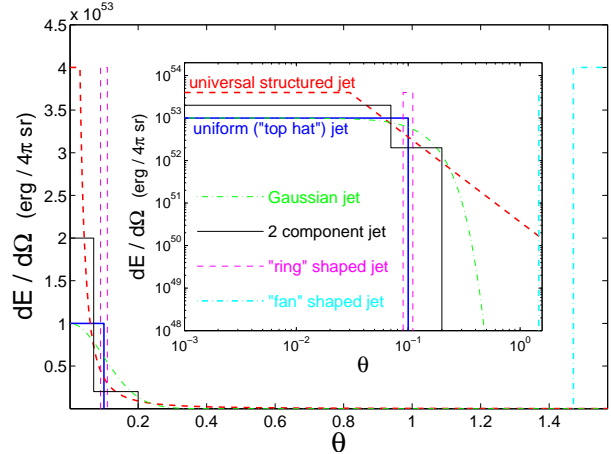}
\caption{An illustration of various jet structures that have been
  discussed in the literature, in terms of the distribution of their
  energy per solid angle, $\epsilon = dE/d\Omega$, with the angle
  $\theta$ from the jet symmetry axis, both in semi-logarithmic scale
  ({\it main figure}) and in log-log scale ({\it big inset}). Both the
  normalization of $dE/d\Omega$ and the typical angular scale may vary
  in most models, and their values shown here were chosen to be more
  or less ``typical''. \citep[from][]{Granot05}.}
\label{fig:jet_structures}
\end{figure}

Other jet structures have also been proposed in the literature. Figure
\ref{fig:jet_structures} illustrates different jet structures that
have been discussed in the literature in terms of their distribution
of $\epsilon(\theta)$. A jet with a Gaussian angular profile
\citep{ZM02,KG03} may be thought of as a more realistic version of a
uniform jet, where the edges are smooth rather than sharp. A Gaussian
$\epsilon(\theta) \propto \exp(-\theta^2/2\theta_c^2)$ is
approximately intermediate between the UJ and USJ models, but it is
closer to the UJ model than to the USJ model with $\epsilon \propto
\theta^{-2}$ in the sense that for a Gaussian $\epsilon(\theta)$ the
energy in the wings of the jet is much smaller than in its core,
whereas for a USJ with $\epsilon \propto \theta^{-2}$ wings there is
equal energy per decade in the wings, and therefore the wings
contain more energy than the core (by about an order of magnitude).

Another jet structure that received some attention recently is a
two-component jet model
\citep{Pederson98,Frail00,Berger03b,Huang04,PKG05,Wu05} with a narrow
uniform jet of initial Lorentz factor $\Gamma_0 \gtrsim 100$
surrounded by a wider uniform jet with $\Gamma_0 \sim 10 -
30$. Theoretical motivation for such a jet structure has been found
both in the context of the cocoon in the collapsar model
\citep{R-RCR02} and in the context of a hydromagnetically driven
neutron-rich jet \citep{VPK03}. This model has been invoked in order
to account for sharp bumps (i.e., fast rebrightening episodes) in the
afterglow light curves of GRB 030329 \citep{Berger03b} and XRF 030723
\citep{Huang04}. A different motivation for proposing this jet
structure is in order to account for the energetics of GRBs and X-ray
flashes and reduce the high efficiency requirements from the prompt
gamma-ray emission \citep{PKG05}.

More ``exotic'' jet structures have also been considered. One example
is a jet with a cross section in the shape of a ``ring,'' sometimes
referred to as a ``hollow cone'' \citep{LE93,LE00,EL03,EL04,LB05},
which is uniform within $\theta_c < \theta < \theta_c + \Delta\theta$
where $\Delta\theta \ll \theta_c$. Another example is a ``fan''- or
``sheet''-shaped jet \citep{Thompson05} where a magnetocentrifugally
launched wind from the proto-neutron star, formed during the supernova
explosion in the massive star progenitor, becomes relativistic as the
density in its immediate vicinity drops and is envisioned to form a
thin sheath of relativistic outflow that is somehow able to penetrate
through the progenitor star along the rotational equator, forming a
relativistic outflow within $\Delta\theta \ll 1$ around $\theta =
\pi/2$ (or $\theta_c = \pi/2 - \Delta\theta/2$).\footnote{This has
been suggested as a possible jet structure within this model, but the
final jet structure is by no means clear, and other jet structures
might also be possible within this model (T. A. Thompson 2005, private
communication).}

\subsection{Statistical Studies}
\label{subsec:statistics}

One approach for constraining the jet structure is through statistical
studies of the prompt emission. In particular, a convenient observable
is the $\log N - \log S$ distribution, where $N$ is the number of GRBs
observed above a limiting peak photon flux $S$
\citep{Firmani04,GPW05,GGB05,Xu05}. In this type of study one needs to
assume both the intrinsic GRB event rate (which is usually assumed to
follow the star formation rate), and the luminosity function which
depends on the jet structure through the isotropic equivalent
luminosity $L(\theta)$ as a function of the angle $\theta$ from the
jet symmetry axis. The latter depends in turn on the angular
distribution of the energy per solid angle in the jet
$\epsilon(\theta)$ and the gamma-ray efficiency
$\epsilon_\gamma(\theta)$, whose product (times $4\pi$) provides the
isotropic equivalent energy output in gamma-rays, $E_{\rm
\gamma,iso}(\theta)$, and the assumed distribution of the peak
isotropic equivalent luminosity for a given $E_{\rm \gamma,iso}$.

\begin{figure}[!t]
\vspace{-0.1cm}
\hspace{-0.27cm}
\includegraphics[width=1.09\columnwidth]{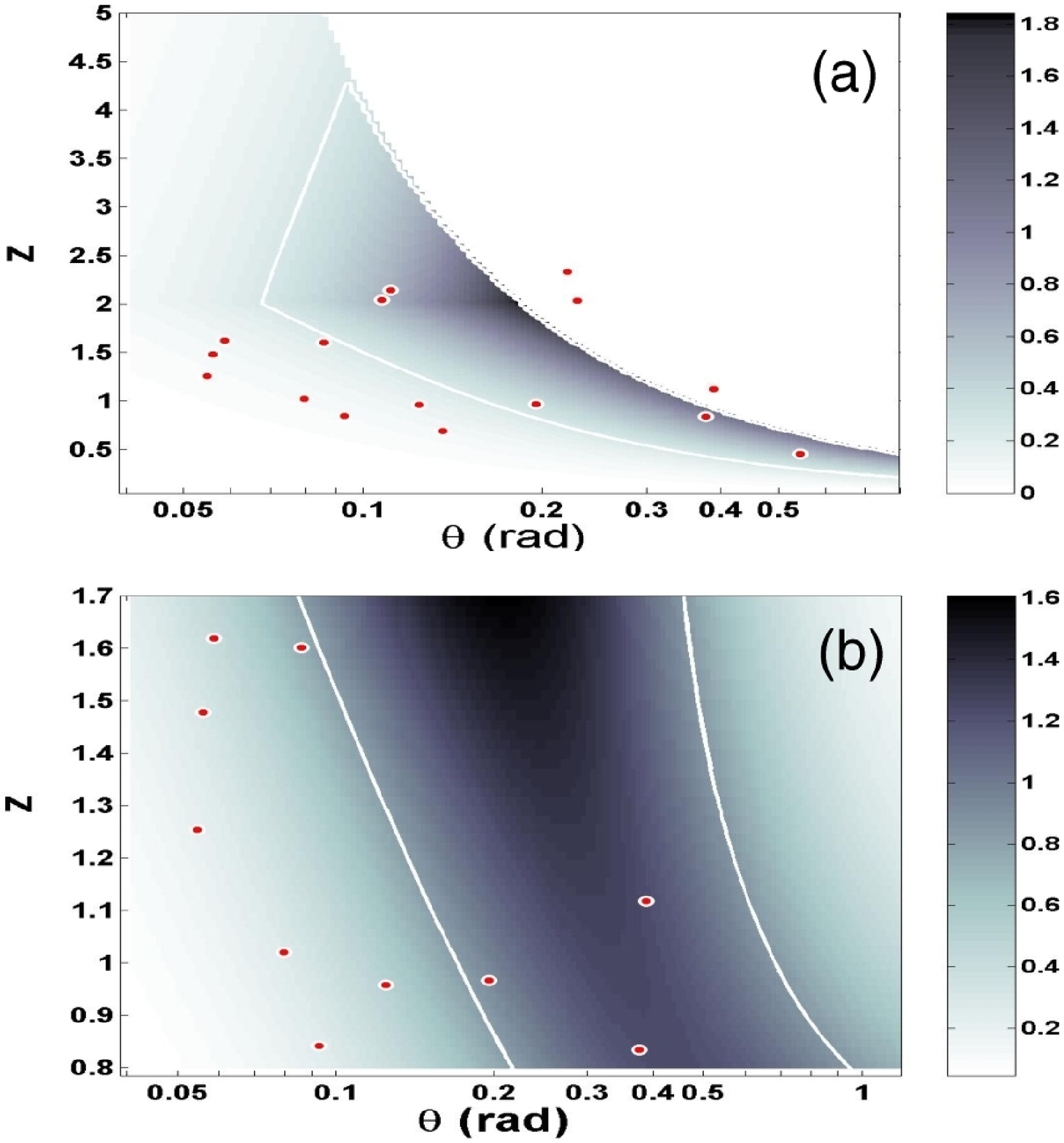}
\caption{The probability distribution, $dN/dz d\ln\theta$, of the
  observed GRB rate as a function of redshift $z$ and viewing angle
  $\theta$, as predicted by the USJ model \citep[from][]{NGG04}. The
  white contour lines confine the minimal area that contains
  $1\;\sigma$ of the total probability. The circles denote the 16 GRBs
  with known $z$ and $\theta$ from the sample of \citet{BFK03}. (a)
  The model parameters are similar to those of \citet{PSF03}. This
  figure is the 2D realization of their Figure 1. (b) Here a limited
  range in redshift is used, $0.8 < z < 1.7$ (containing 10 out of the
  16 data points), in order to minimize redshift selection effects and
  reduce the sensitivity of the results to the unknown GRB
  rate. Measurement errors of 20\% in $\ln\theta$ ($\sigma_{\ln\theta}
  = 0.2$) were included and a log-normal distribution in the effective
  duration that they deduced from observations.}
\label{fig:dNdzdth}
\end{figure}

For the USJ model $\epsilon(\theta)$ and $\epsilon_\gamma(\theta)$ are
assumed to be power laws in $\theta$ and one needs to specify their
power law indexes, as well as the jet core angle $\theta_c$ and outer
edge $\theta_{\rm max}$ \citep{GGB05}. For the UJ model one needs to
specify the distribution of jet half-opening angles $\theta_0$, which
can be taken to be a power law distribution in the range $\theta_{\rm
min} < \theta_0 < \theta_{\rm max}$. These simple forms of the USJ and
UJ models are degenerate, since they both produce a power law
luminosity function \citep{GGB05}, which provides an adequate fit to
the data. The total rate of energy release in gamma-rays must be the
same in both models (and match the observed rate), where in the USJ
model it is released in fewer more energetic events \citep[by a factor
of $\sim 10$; see equation 14 of][]{GGB05}, while in the UJ model it
is released in more numerous less energetic events (i.e. the UJ model
predicts a larger intrinsic event rate).

Another statistical approach for constraining the jet structure is
through the distribution of the observed number of GRBs $N$ as a
function of the angle $\theta$ that is inferred from the observed jet
break times in the afterglow light curves, where $\theta$ corresponds
to the jet half-opening angle $\theta_0$ in the UJ model and to the
viewing angle $\theta_{\rm obs}$ in the USJ model.  The observed
$dN/d\theta$ distribution agrees reasonably well with the predictions
of the USJ model \citep{PSF03}, which had been argued to support the
USJ model (since in the competing UJ model there is an additional
freedom in the choice of the probability distribution for $\theta_0$
which would make it easier to fit these observations). However, when
the known redshifts $z$ of the GRBs in the same sample are also taken
into account, then the predictions of the USJ model for the two
dimensional distribution of observed GRBs with $\theta$ and $z$,
$dN/dzd\theta$, is found to be in very poor agreement with
observations \citep{NGG04}. This can be best seen for a relatively
narrow range in $z$ (see lower panel of Figure \ref{fig:dNdzdth}), in
which the USJ model predicts that most GRBs should be near the upper
end of the observed range in $\theta$, while in the observed sample
most GRBs are near the lower end of that range. Since the available
sample was very inhomogeneous (i.e., involved many different
detectors), it should be taken with care and cannot be used to rule
out the USJ model. Nevertheless, it strongly disfavors the USJ model.

\subsection{Linear Polarization}
\label{subsec:lin_pol}

\begin{figure}[!t]
\vspace{-0.60cm}
\includegraphics[width=1.09\columnwidth]{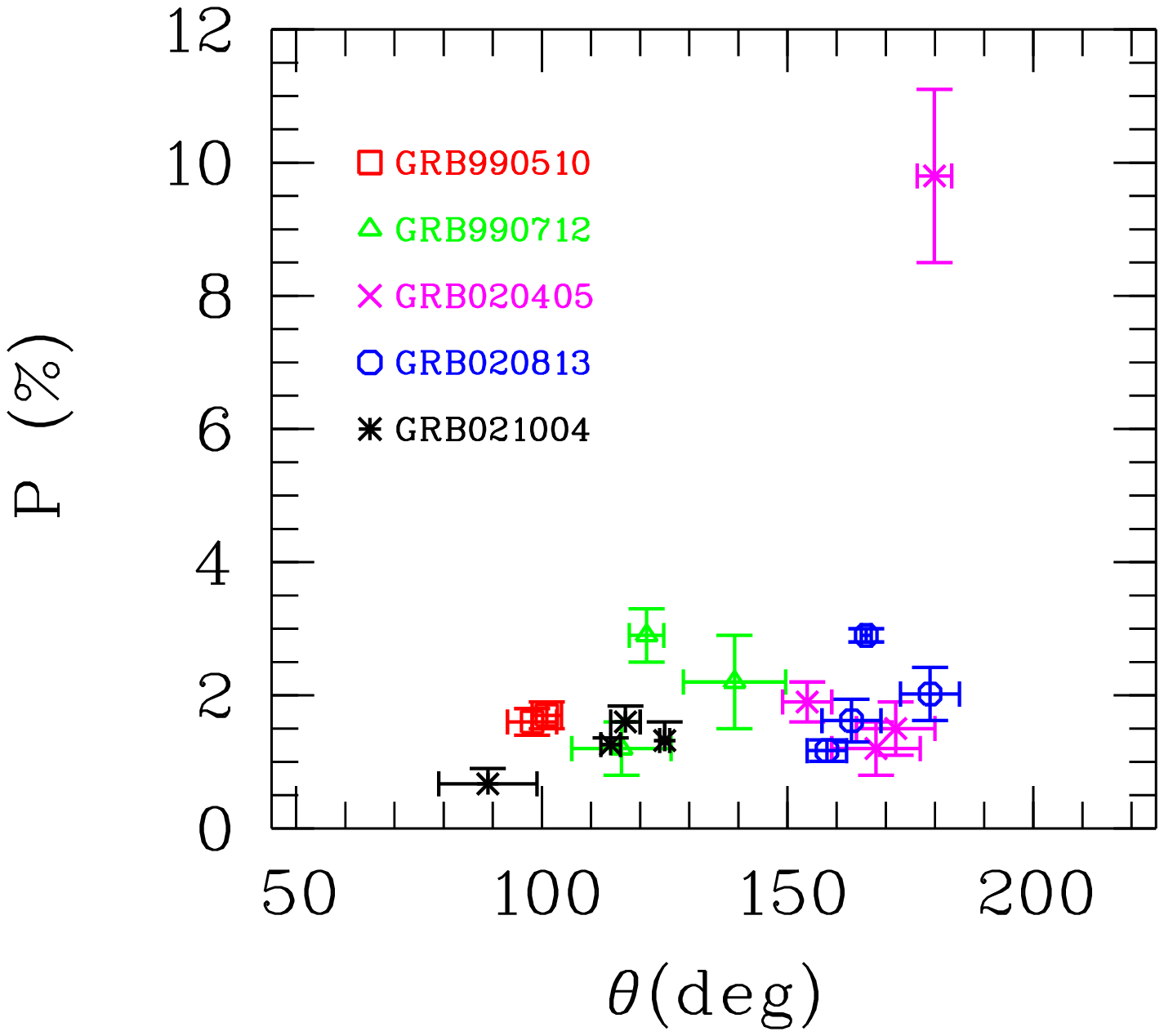}
\\
\vspace{-0.8cm}
\\
\includegraphics[width=1.09\columnwidth]{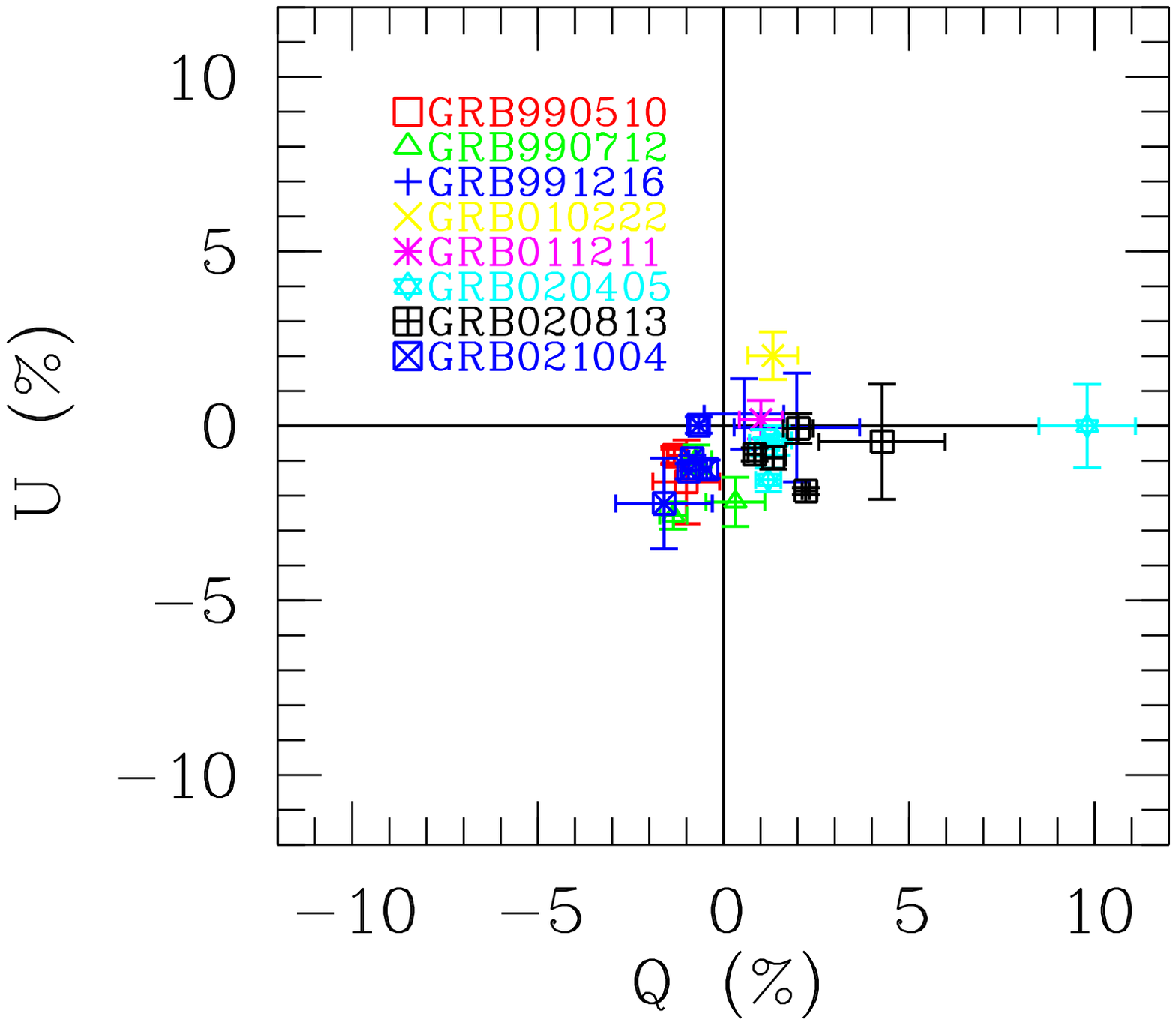}
\vspace{-1.6cm}
\\
\caption{Summary of afterglow polarization measurement up to 2002
\citep[from][]{Covino04}. {\it Top panel}: degree of polarization and
position angle for all the positive detections, i.e. upper limits are
excluded. {\it Bottom panel}: Q and U Stokes parameters for all the
available data, i.e. including upper limits.}
\label{fig:AG_pol}
\end{figure}

Linear polarization at the level of a few percent has been detected in
the optical or NIR afterglow of several GRBs
\citep{Covino99,Wijers99,Rol00,Covino03}, as is illustrated in Figure
\ref{fig:AG_pol}. This was considered as a confirmation that
synchrotron radiation is the dominant emission mechanism in the
afterglow. The most popular explanation for the observed linear
polarization had been synchrotron emission from a jet
\citep{Sari99,GL99}. In this model the magnetic field is produced at
the afterglow shock and possesses axial symmetry about the shock
normal. In this picture there would be no net polarization for a
spherical outflow, since the polarization from the different parts of
the afterglow image would cancel out, and a jet geometry together with
a line of sight that is not along the jet axis (but still within the
jet aperture, in order to see the prompt GRB) is needed in order to
break the symmetry of the afterglow image around our line of sight.

For a uniform jet (the UJ model) this predicts two peaks in the
polarization light curve around the jet break time $t_j$, if
$\Gamma\theta_j < 1$ decreases with time at $t > t_j$
\citep{GL99,Rossi04}, or even three peaks if $\Gamma\theta_j \approx
1$ at $t> t_j$ \citep{Sari99}, where in both cases the polarization
vanishes and reappears rotated by $90^\circ$ between adjacent peaks.
The latter is a distinct signature of this model. For a structured jet
(the USJ model), the polarization position angle is expected to remain
constant in time, while the degree of polarization peaks near the jet
break time $t_j$ \citep{Rossi04}. A similar qualitative behavior
is also expected for a Gaussian jet, or other jet structures with a
bright core and dimmer wings (although there are obviously some
quantitative differences).

The different predictions for the afterglow polarization light curves
for different jet structures raise the hopes that afterglow
polarization observations may constrain the jet structure. In
practice, however, the situation is much more complicated, mainly
since the observed polarization depends not only on the jet geometry,
but also on the magnetic field configuration in the emitting region,
which is not known very well. For example, an ordered magnetic field
component in the emitting region (e.g. due to a small ordered magnetic
field in the external medium) may dominate the polarized flux, and
therefore the polarization light curves, even if it is sub-dominant in
the emitting region compared to a random (shock generated) magnetic
field component in terms of the total energy in the magnetic field
\citep{GK03}.  Other models for afterglow polarization include a
magnetic field that is coherent over patches of a size comparable to
that of causally connected regions \citep{GW99}, and polarization that
is induced by microlensing \citep{LP98} or by scintillations in the
radio \citep{ML99}.

\begin{figure}[!t]
\vspace{0.14cm}
\includegraphics[width=0.90\columnwidth]{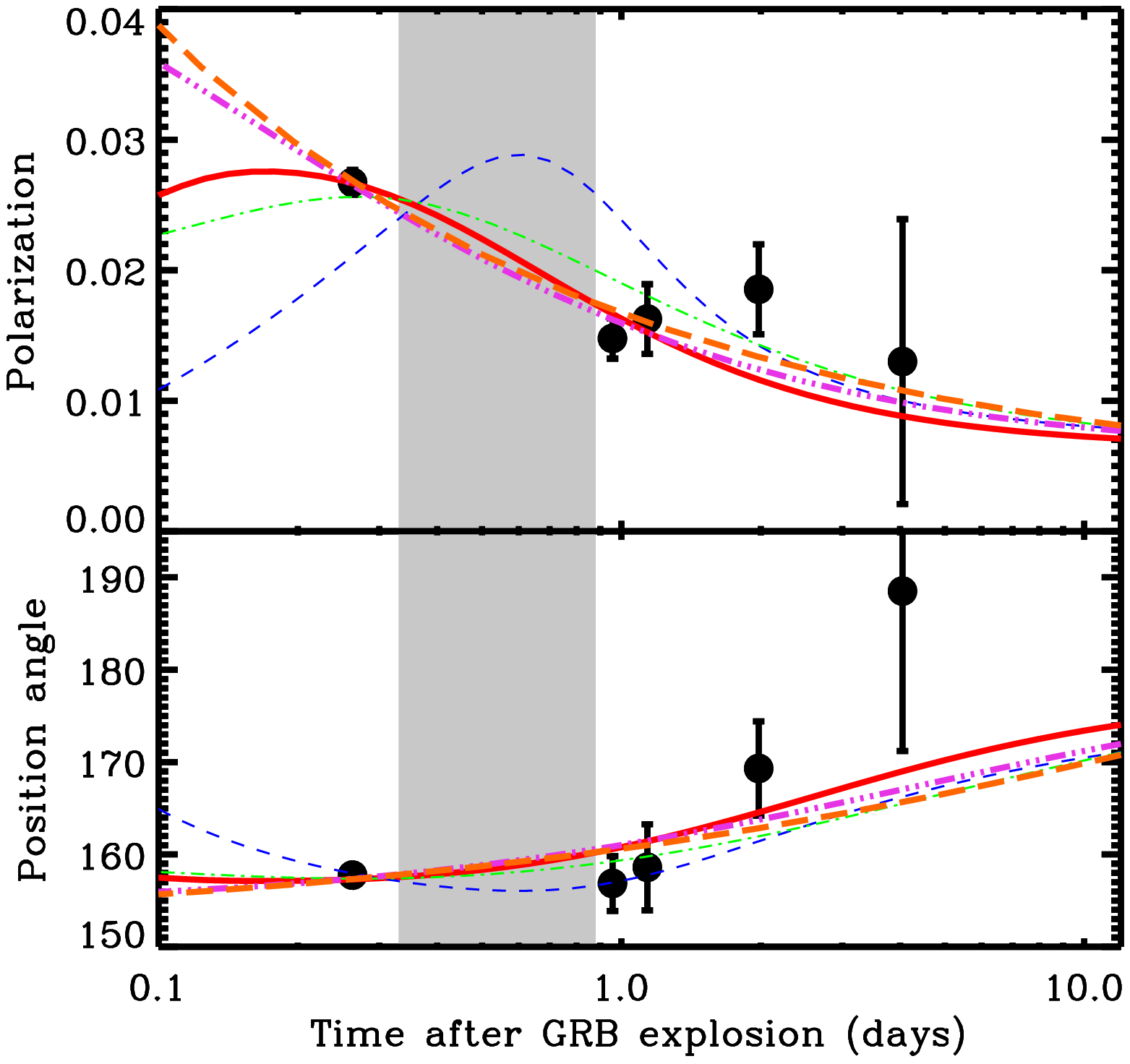}
\\
\\
\includegraphics[width=0.90\columnwidth]{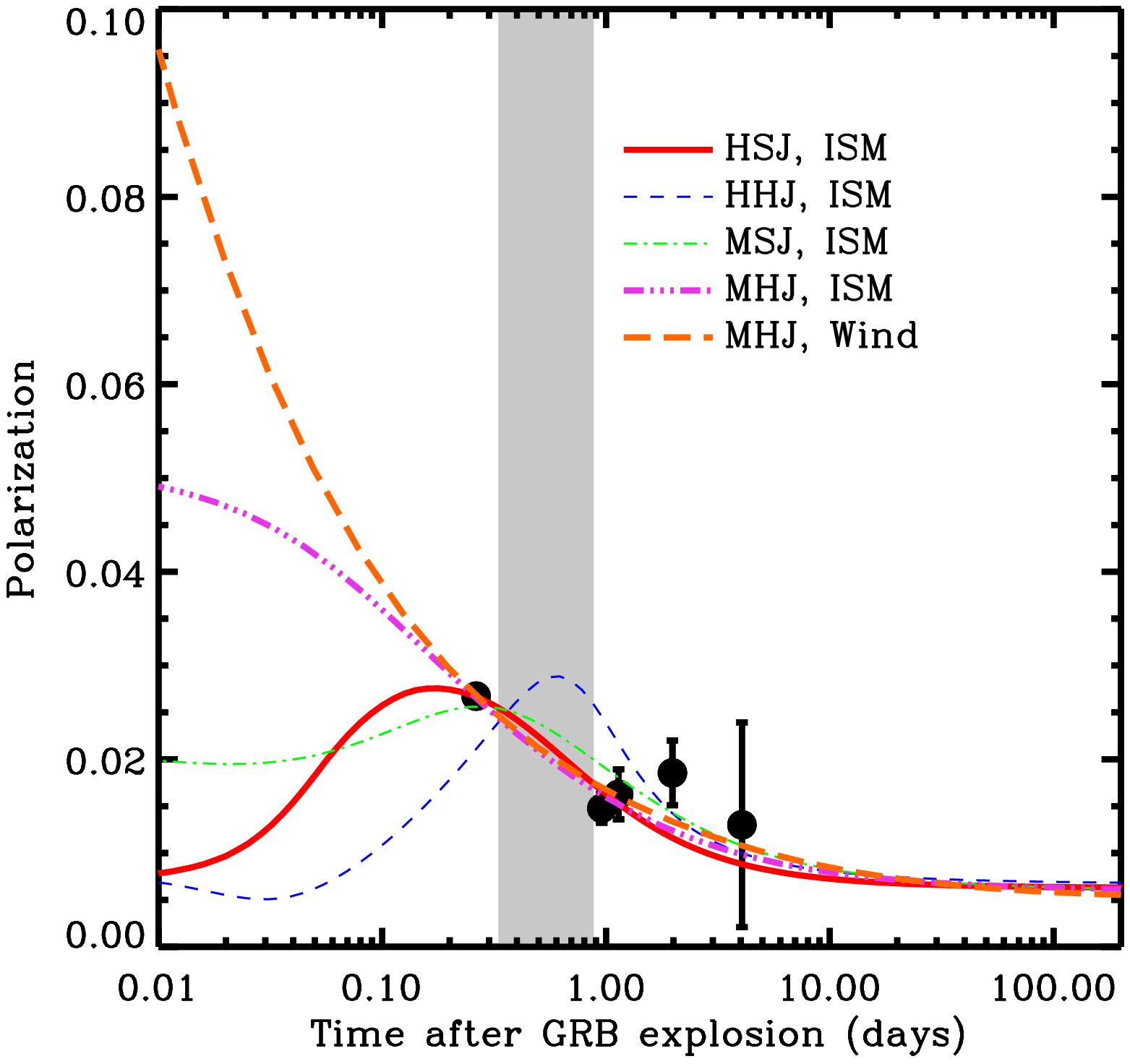}
\\
\vspace{-0.6cm}
\\
\caption{Fits of the predicted polarization for different jet
  structures and magnetic field configuration to the optical afterglow
  polarization data for GRB~020813 \citep[from][]{Lazzati04}. The {\it
  upper panel} shows the degree of polarization and position angle
  around the jet break time, which is also where the observations are
  concentrated, while the {\it bottom panel} shows the degree of
  polarization over a wider range in time and demonstrates that the
  differences between the various models are more pronounced at $t
  \lesssim t_j$. The shaded region denotes the allowed range for the
  jet break time $t_j$. The external density is taken to be either
  uniform (ISM, $k = 0$) or a stellar wind (Wind, $k = 2$). The
  different models are labeled by a three letter acronym where the
  first letter describes the magnetic field (`H' is for hydrodynamic,
  i.e. sock produced magnetic field, while `M' is for magnetized,
  i.e. with an ordered toroidal magnetic field) while the second
  letter describes the jet structure (`S' for a structured jet, and
  `H' for a homogeneous or uniform jet).}
\label{fig:AG_pol_fit}
\end{figure}

An additional complication arises in afterglow light curves that
exhibit variability, since a variable afterglow light curve is
expected to be accompanied by a variable polarization light curve,
both in the degree of polarization and in its position angle
\citep{GK03}. This prediction has been confirmed in GRB~021004
\citep{Rol03}, where it had been interpreted both in the context of
angular inhomogeneities within the jet \citep[i.e. a ``patchy
shell'',][]{NO04} and as discrete episodes of energy injection into
the afterglow shock \citep[i.e. ``refreshed shocks'',][]{BGJ04}, and
later also in GRB~030329 \citep{Greiner03}. 

Perhaps the best monitored polarization light curve of a smooth
afterglow, which does not suffer from the complications mentioned
above, is GRB~020813 \citep{Gorosabel04} where the polarization
position angle is roughly constant while the degree of polarization
decreased by a factor of $\sim 2$ from $\sim 0.5t_j$ to $\sim
2t_j$. The constant position angle across the jet break time $t_j$
disfavors a uniform jet with a shock generated magnetic field
\citep{Sari99,GL99}, as well as patches of uniform field \citep{GW99}
where the position angle (as well as the degree of polarization) is
expected to change randomly on time scales $\Delta t \lesssim t$.

\citet{Lazzati04} have contrasted different models for the jet
structure and magnetic field configuration with the polarization data
for GRB~020813 (see Figure \ref{fig:AG_pol_fit}), and concluded that
the data support either (i) a structured jet (USJ) or a jet structure
where most of the jet energy is in a narrow core while its wings
contain less energy (such as a Gaussian jet) with a shock produced
magnetic field (where the field is not purely in the plane of the
shock but still possesses significant anisotropy) or (ii) a uniform
jet\footnote{\citet{Lazzati04} find that a structured jet (USJ) with
an ordered toroidal field component that dominates the polarization is
still hard to rule out, even though it provides a poor fit to the
polarization data of GRB~020813, due to the model uncertainties
regarding the mixing of such an ordered field component with the shock
generated random field.} with an ordered toroidal magnetic field
component which dominates the polarized flux together with a random
magnetic field component that dominates the total flux (and the total
magnetic energy) in order for the polarization not to exceed the
observed value.

We conclude that while the afterglow polarization light curves may
provide useful constraints on the jet structure and the magnetic field
configuration in the emitting region, it is in practice rather
difficult to constrain each of these two ingredients separately. That
is, in order to obtain tight constraints on the jet structure, strong
assumptions must be made on the magnetic field configuration, and vice
versa. Nevertheless, as discussed above, interesting constraints have
already been derived from existing data.

\subsection{Afterglow Light Curves}
\label{subsec:afterglow_LC}

\begin{figure}[!t]
\includegraphics[width=1.00\columnwidth]{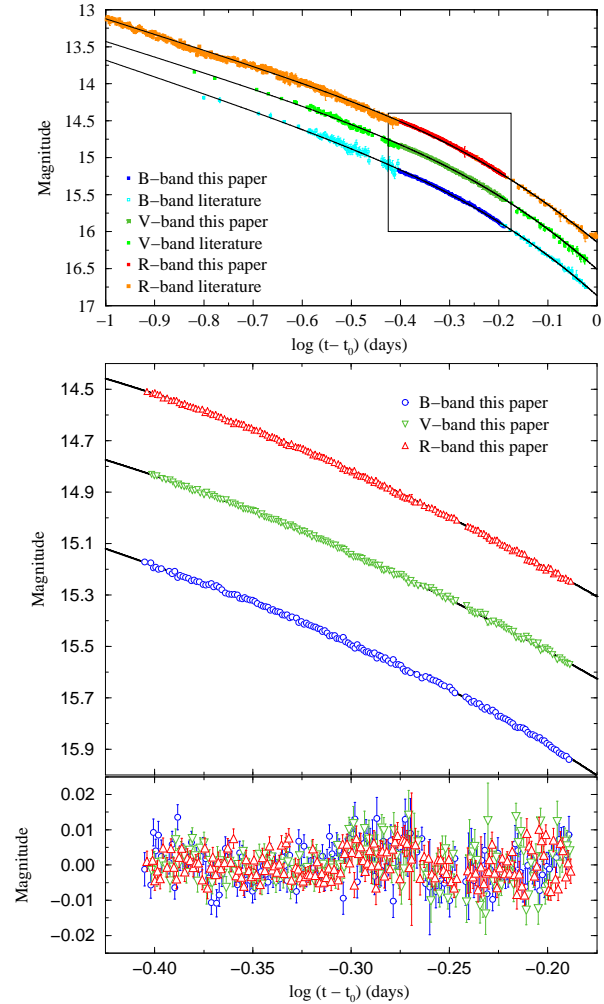}
\caption{The jet break of GRB~030329 at three different optical bands,
  with exquisite temporal sampling \citep[from][]{Gorosabel06}. The
  {\it upper panel} shows a wider time span complemented by data from
  other groups, while the {\it middle panel} shows the data from
  \citet{Gorosabel06}, and the {\it lower panel} shows the residuals
  relative to a fit to a smooth jet break model. The steepening in the
  temporal decay during the jet break is smooth and achromatic.}
\label{fig:GRB030329_jet_break}
\end{figure}

The shape of the afterglow light curves is an important and relatively
robust diagnostic tool for constraining the jet structure. The
afterglow light curves (at least starting from a few hours after the
GRB) are typically described by an initial power law flux decay $F_\nu
\propto t^{-\alpha}$ with $0.7 \lesssim \alpha_1 \lesssim 1.5$ which
steepens into a sharper power law decay ($1.6 \lesssim \alpha_2
\lesssim 2.8$) at the jet break time $t_j$ \citep{ZKK06}. Figure
\ref{fig:GRB030329_jet_break} shows an example of the very well
monitored jet break of GRB~030329. The jet break in the light curve is
usually rather sharp (most of the steepening occurs within a factor of
a few in time) and the increase in the temporal decay index,
$\Delta\alpha = \alpha_2 - \alpha_1$, is typically in the range $0.7
\lesssim \Delta\alpha \lesssim 1.4$ \citep{ZKK06}. Different jet
structures may be tested by their ability to reproduce these observed
properties.  Below we describe the resulting constraints for various
jet models.

\begin{figure}[!t]
\includegraphics[width=1.00\columnwidth]{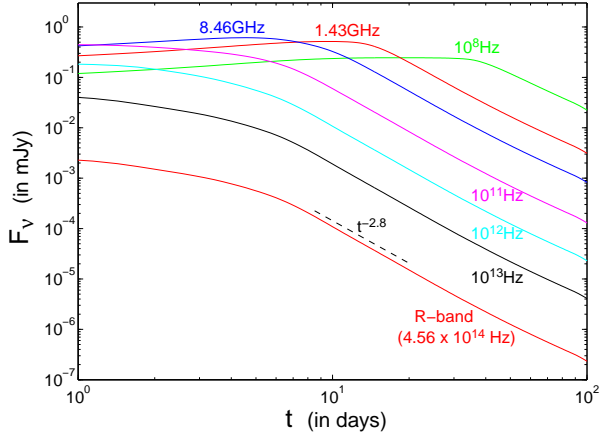}
\caption{Afterglow light curves for an initially uniform jet whose
  evolution is calculated using a hydrodynamic simulation, at
  different observe frequencies, for an observer along the jet
  symmetry axis \citep[from][]{Granot01}.}
\label{fig:UJ_LC}
\end{figure}

{\bf Uniform Jet:} Figure \ref{fig:UJ_LC} shows the afterglow light
curves for an initially uniform jet whose evolution is calculated
using a hydrodynamic simulation \citep{Granot01}. The initial
conditions are a cone of half-opening angle $\theta_0$ taken out of
the spherical self-similar \citet{BM76} solution (see \S
\ref{subsec:sim}). The shape of the light curves is nicely consistent
with those observed in real afterglows, particularly in terms of the
sharpness of the jet break, where the observed diversity can be
attributed to different viewing angles within the jet aperture,
$\theta_{\rm obs} < \theta_0$ (see upper panel of
Figure \ref{fig:off-axis_LC}).

{\bf Gaussian Jet:} the afterglow light curves for a Gaussian jet
where $\epsilon(\theta) = \epsilon_0\exp(-\theta^2/2\theta_c^2)$, that
is observed at viewing angles inside the core of the jet ($\theta_{\rm
obs} < \theta_c$) are rather similar to those for a uniform jet
\citep{KG03,Rossi04,GR-RP05}, and reasonably consistent with afterglow
observations.

\begin{figure}[!t]
\vspace{-0.1cm}
\includegraphics[width=1.00\columnwidth]{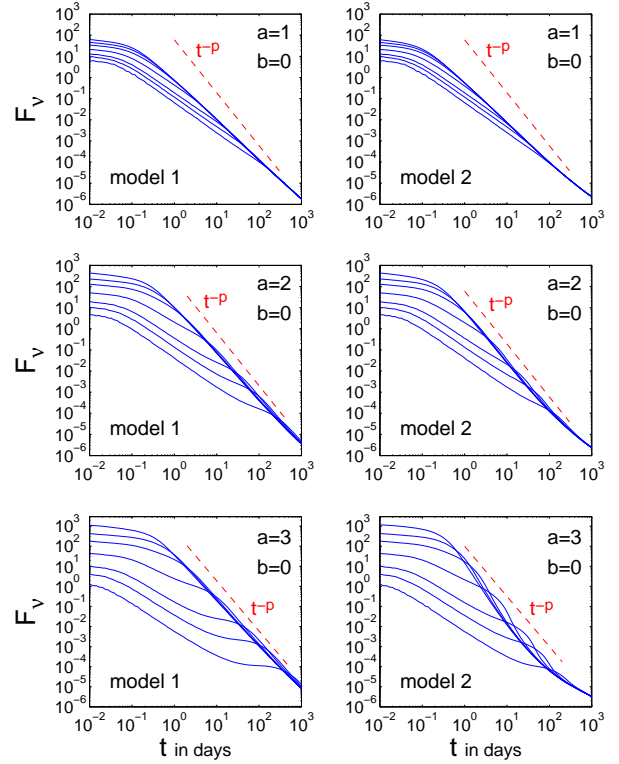}
\caption{Afterglow light curves for a structured jet where initially
$\epsilon \propto \theta^{-a}$ and $\Gamma \propto \theta^{-b}$
outside some core angle $\theta_c$ \citep[from][]{GKu03}. The
different curves, from top to bottom, are for viewing angles
$\theta_{\rm obs} = 0.01,\,0.03,\,0.05,\,0.1,\,0.2,\,0.3,\,0.5$, where
$\theta_c = 0.02$, $p = 2.5$, $\epsilon_e = \epsilon_B = 0.1$, $n_{\rm
ext} = 1\;{\rm cm^{-3}}$, $\Gamma(\theta = 0,t_0) = 10^3$, and the
total energy in the jet is $10^{52}\;$erg.  In model~1
$\epsilon(\theta)$ does not change with time. In model~2,
$\epsilon(\theta,t)$ evolves such that it is proportional to the
average over its initial distribution, $\epsilon(\theta,t_0)$, over
the range in $\theta$ out to which a sound wave could propagate from
$t_0$ up to $t$ \citep[see][for details]{GKu03}.}
\label{fig:USJ_a}
\end{figure}

{\bf Structured Jet:} one can consider a jet with a narrow core and
wings where both $\epsilon$ and $\Gamma$ are initially power laws in
the angle $\theta$ from the jet symmetry axis: $\epsilon \propto
\theta^{-a}$ and $\Gamma_0 \propto \theta^{-b}$, outside some (narrow)
core angle $\theta_c$. A comparison between the resulting afterglow
light curves and afterglow observations can then be used to constrain
the power law indexes $a$ and $b$. Such an analysis \citep{GKu03}
suggests that $0 \lesssim b \lesssim 1$ and $a \approx 2$ (or $1.5
\lesssim a \lesssim 2.5$, see Figure \ref{fig:USJ_a}). 

The upper limit on $b$ comes from the fact that a large value of $b$
implies a small initial Lorentz factor at large viewing angles,
$\Gamma_0(\theta_{\rm obs}\gg\theta_c)$, since it is hard for
$\Gamma_0(\theta = 0)$ to exceed $10^4$. A small initial Lorentz
factor along the line of sight at large viewing angles would result in
a large decelerations time along the line of sight and therefore an
initially rising light curve, up to relatively late times, which is
not seen in observations. Furthermore $\Gamma_0(\theta_{\rm obs})
\gtrsim 100$ is needed in order to produce the prompt gamma-ray
emission. Figure \ref{fig:USJ_a} shows light curves for $b = 0$ and $a
= 1,\,2,\,3$ and different viewing angles. For $a = 1$ the change in
the temporal decay index across the jet break, $\Delta\alpha$, is too
small (compared to its observed values) and the post-jet break decay
slope is not steep enough. For $a = 3$ there is either a very
pronounced flattening in the light curve before the jet break time or
the temporal decay slope after the jet break is extremely steep
(neither of which is seen in afterglow observations). This suggests $a
\approx 2$ (or $1.5 \lesssim a \lesssim 2.5$).

\begin{figure}[!t]
\includegraphics[width=1.00\columnwidth]{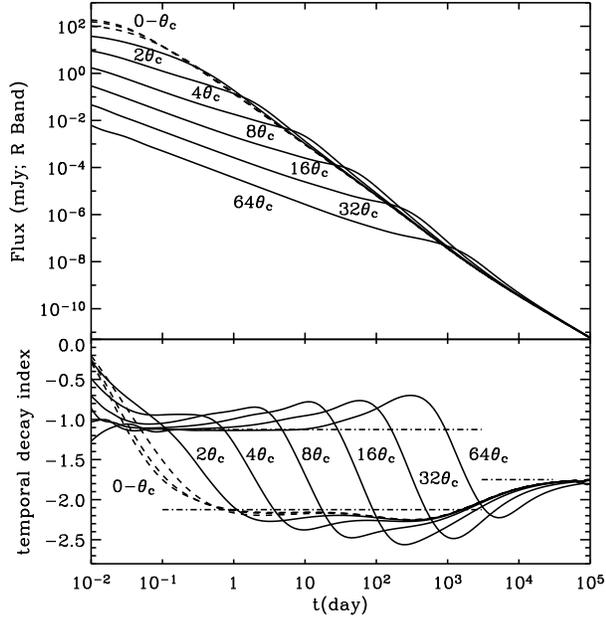}
\caption{Optical (R-band) light curves ({\it upper panel}) and the
  temporal decay index ({\it lower panel}) for a structured jet with
  no lateral spreading \citep[from][]{Rossi04}, for $a = 2$, $b = 0$,
  $E_{\rm k,iso}(\theta = 0) = 2\times 10^{54}\;$erg, $n_{\rm ext} =
  1\;{\rm cm^{-3}}$, $\epsilon_e = 0.01$, $\epsilon_B = 0.005$,
  $\theta_c = 1^\circ$. The different curves are for different viewing
  angles, $\theta_{\rm obs}$, which are labeled in units of the jet
  core angle, $\theta_c$.}
\label{fig:Rossi04}
\end{figure}

As can be seen from Figure \ref{fig:Rossi04}, even for $a = 2$ and $b
= 0$ there is a flattening in the light curve before the jet break
time, which becomes more pronounced at large viewing angles
($\theta_{\rm obs} \gg \theta_c$). This arises since the bright core
of the jet becomes visible, while the value of $\epsilon$ along the
line of sight is much smaller in comparison for large viewing angles,
which more than compensates for the relatively small fraction of the
visible region that is occupied by the jet core. That is, the average
energy per solid angle in the observed region initially increases with
time until the core of the jet becomes visible, and then decreases
with time. Such a flattening has not been observed so for, but since
it is most pronounced at large viewing angles for which the jet break
time is large and the flux around that time is low, the observations
in this parameter range are usually rather sparse, making it hard to
rule out this model on these grounds.

{\bf Two Component Jet Model:} the light curves for this jet structure
have been calculated analytically \citep{PKG05} or semi-analytically
\citep{Huang04,Wu05}, and it has been suggested that this model can
account for sharp bumps (i.e., fast rebrightening episodes) in the
afterglow light curves of GRB 030329 \citep{Berger03b} and XRF 030723
\citep{Huang04}. It has been later demonstrated, however, that effects
such as the modest degree of lateral expansion that is expected in
impulsive relativistic jets and the gradual hydrodynamic transition at
the deceleration epoch smoothen the resulting features in the
afterglow light curve, so that they cannot produce features as sharp
as those mentioned above \citep{Granot05}. One of the motivations for
a two component jet was to alleviate the efficiency requirements on
the prompt gamma-ray emission \citep{PKG05} if the wide jet dominates
the total energy and late time afterglow emission, while the narrow
jet is responsible for the prompt gamma-ray emission. However, the
more recent {\it Swift} observations, which show a flat decay phase in
the early X-ray afterglow \citep{Nousek06}, are inconsistent which
this picture and suggest a high gamma-ray efficiency also for this jet
model under the standard assumptions of afterglow theory
\citep{GKP06}.

\begin{figure}[!t]
\vspace{-0.05cm}
\includegraphics[width=1.00\columnwidth]{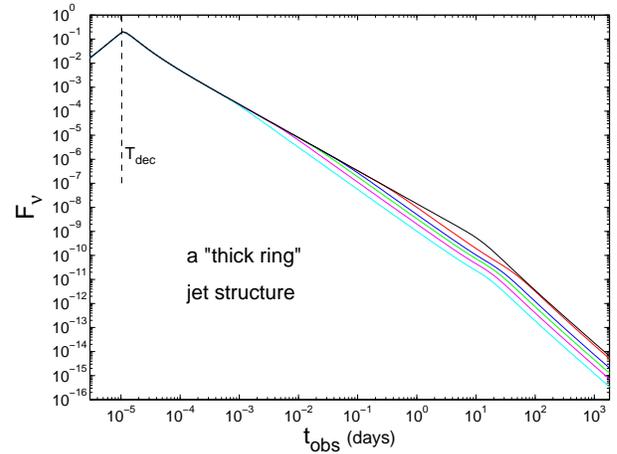}
\caption{Light curves for a jet with an angular structure of a ring,
  where $\epsilon$ is uniform for $\theta_c < \theta < \theta_c +
  \Delta\theta$, and sharply drops outside of this range in $\theta$,
  for various fractional widths, viewed from within the jet. The upper
  line is for a uniform jet viewed from along its symmetry axis
  ($\theta_c = \theta_{\rm obs} = 0$, $\Delta\theta = 0.2$) and is
  included for comparison, while the other lines are for a ring shaped
  jet with $\theta_c = 0.1$ and $\theta_c/\Delta\theta =
  1,\,2,\,3,\,5,\,10$, from top to bottom, viewed from $\theta_{\rm
  obs} = \theta_c + \Delta\theta/2$. A uniform external density ($k =
  0$), $p = 2.5$, and no lateral expansion are assumed
  \citep[from][]{Granot05}.}
\label{fig:ring}
\end{figure}

{\bf Ring Shaped Jet:} Figure \ref{fig:ring} shows light curves for a
ring shaped jet, viewed from within the emitting region, for different
fractional width of the ring \citep{Granot05}. For a thin ring, whose
width is smaller than its radius, $\Delta\theta < \theta_c$, the jet
break is divided into two distinct and smaller breaks, the first
occurring when $\Gamma\Delta\theta \sim 1-2$ (i.e. when the width of
the jet becomes visible) and the second when $\Gamma\theta_c \sim 1/2$
(i.e. when the whole jet becomes visible). This is inconsistent with
the large steepening that is observed across the jet break, and
suggests a ``thick'' ring, with $\Delta\theta \gtrsim \theta_c$. Even
for such a thick ring with $\Delta\theta = \theta_c$ the jet break is
still not quite sharp enough to match observations according to the
model shown in Figure \ref{fig:ring}, however, the inclusion of some
lateral expansion which would tend to smoothen the edges of the jet
and fill in the center of the thick ring would probably be enough to
make the light curve for $\Delta\theta \gtrsim \theta_c$ consistent
with observed afterglow light curves. The fact that the beaming cone
extends out to an angle of $\sim \Gamma^{-1}$ from the edge of the jet
further contributes to smoothing the jet break for a thick ring jet,
and contributes to the observed flux from lines of sight near the edge
of the jet \citep{Eichler05}.

\begin{figure}[!t]
\vspace{-0.05cm}
\includegraphics[width=1.00\columnwidth]{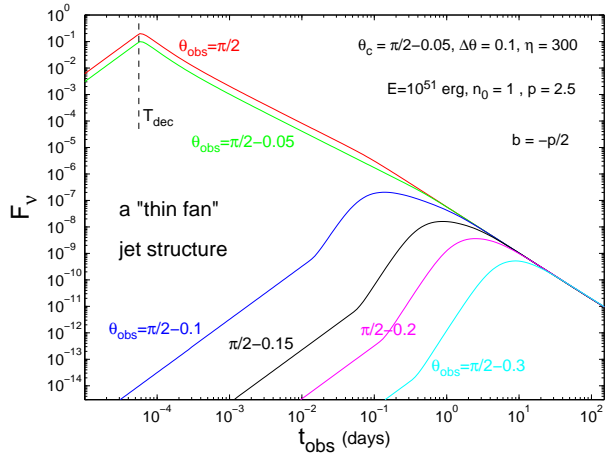}
\caption{Light curves for a jet with an angular structure of a thin
 fan, with an opening angle of $\Delta\theta = 0.1$ centered on
 $\theta = \pi/2$, i.e. $\theta_c = \pi/2 - \Delta\theta /2 = \pi/2 -
 0.05$ \citep[from][]{Granot05}.}
\label{fig:fan}
\end{figure}

{\bf Fan Shaped Jet:} Figure \ref{fig:fan} shows the light curves for
a jet in the shape of a thin fan, for different viewing angles. This
jet structure is a limiting case of the ring shaped jet where
$\theta_c = \pi/2 - \Delta\theta/2$, and the jet occupies an angle of
$\Delta\theta \ll 1$ centered around $\theta = \pi/2$.  In this case
the second jet break from the thin ring jet that had been discussed
above occurs only when the jet becomes sub-relativistic, and is
indistinguishable from the non-relativistic transition. This leaves
only one jet break, when $\Gamma\Delta\theta \sim 1-2$ and the width
of the jet becomes visible. The steepening in the flux decay rate
across this jet break is at most half of that for a uniform jet
(exactly half with no lateral expansion, and slightly less than half
with rapid lateral expansion), and is too small compared to
observations \citep{Granot05}. This practically rules out a jet in the
shape of a thin fan.

\subsection{Off-Axis Viewing Angles}
\label{subsec:off-axis}

\begin{figure}[!t]
\includegraphics[width=1.00\columnwidth]{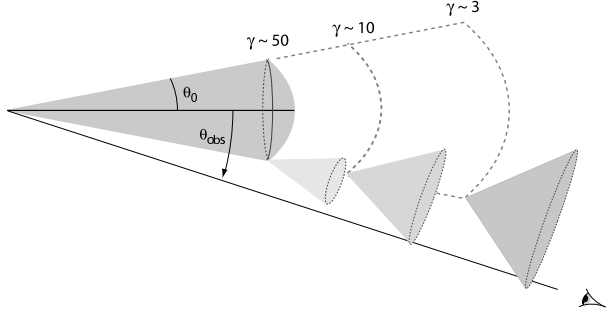}
\caption{An illustrative diagram of the emission from a uniform
relativistic jet with sharp edges and half-opening angle $\theta_0$,
that is seen by an off-axis observer whose line of sight makes an
angle $\theta_{\rm obs} > \theta_0$ with the jet axis. Because of
relativistic beaming (i.e. aberration of light) the emission from each
part of the jet is beamed into a narrow cone of half-opening angle
$\gamma^{-1}$ around its direction of motion in the observer
frame. During the prompt emission (and the very early afterglow) the
Lorentz factor of the jet is large ($\gamma \gtrsim 50$) and therefore
most of the radiation is strongly beamed away from the line of
sight. In this case, the little radiation that is observed comes
mainly from near the edge of the jet, at the point closest to the line
of sight. As the jet decelerates $\gamma$ decreases with time and the
beaming cone grows progressively wider, causing the radiation to be
less strongly beamed, resulting in a rising light curve. The light
curve peaks when $\gamma$ drops to $\sim (\theta_{\rm obs} -
\theta_0)^{-1}$ as the line of sight enters the beaming cone of the
emitting material at the edge of the jet (the middle beaming cone in
the figure), and subsequently decays with time, asymptotically
approaching the light curve for an on-axis observer ($\theta_{\rm obs}
< \theta_0$) at later times. \citep[from][]{GR-RP05}.}
\label{fig:off-axis_sketch}
\end{figure}

When the jet has relatively sharp edges, then the observed emission
from viewing angles outside of the jet aperture is very different than
that from viewing angles within the jet aperture
\citep{PMR98,PM99,MSB00,Granot01,Granot02,DGP02}. If the line of sight
is at an angle of $\Delta\theta_{\rm obs}$ outside the edge of the jet
(i.e. from the nearest point in the jet with significant emission; for
a uniform jet $\Delta\theta_{\rm obs} = \theta_{\rm obs} - \theta_j$)
then at early times when $\Gamma\Delta\theta_{\rm obs} \gg 1$ the
emitted radiation is strongly beamed away from the observer. This is
since the emission is roughly isotropic in the comoving frame of the
emitting material in the jet, and it is collimated within an angle of
$\Gamma^{-1}$ around its direction of motion in the lab frame, due to
relativistic beaming (i.e. aberration of light). As the jet
decelerates, the beaming of radiation away from the line of sight
becomes weaker, resulting in a rising flux. Eventually, when
$\Gamma\Delta\theta_{\rm obs}$ decreases to $\sim 1$, the beaming cone
of the jet encompasses the line of sight (see Figure
\ref{fig:off-axis_sketch}) and the observed flux peaks and starts to
decay, asymptotically joining the decaying light curve for viewing
angles within the initial jet aperture.

\begin{figure}[!t]
\includegraphics[width=1.00\columnwidth]{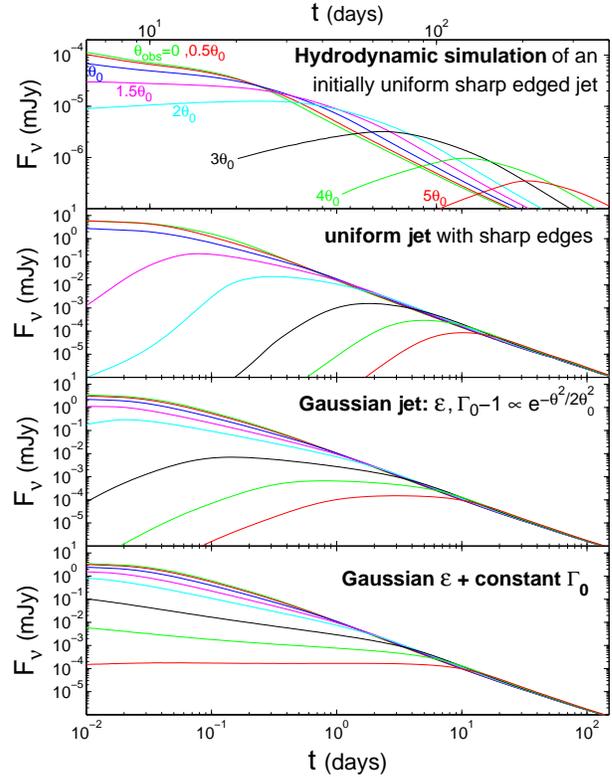}
\caption{Afterglow light curves for different jet structures,
dynamics, and viewing angles. The top panel is from an initially
uniform jet with sharp edges whose evolution is calculated using a
hydrodynamic simulation \citep[taken from Figure 2 of][]{Granot02}. The
remaining three panels are taken from Figure 5 of \citet{Granot05},
where a simplified jet dynamics with no lateral expansion is used. The
second panel is for a uniform jet with sharp edges. The two bottom
panels are for a Gaussian jet, in energy per solid angle, and either a
Gaussian or a uniform initial Lorentz factor. The viewing angles are
$\theta_{\rm obs}/\theta_0 = 0,\,0.5,\,1,\,1.5,\,2,\,3,\,4,\,5$, where
$\theta_0$ is the (initial) half-opening angle for the uniform jet
(two top panels) and the core angle ($\theta_c$) for the Gaussian jet
(two bottom panels). \citep[from][]{EG06}.}
\label{fig:off-axis_LC}
\end{figure}

The afterglow light curves for different jet structures, dynamics, and
viewing angles are shown in Figure \ref{fig:off-axis_LC}. For an
initially uniform jet with sharp edges whose evolution is calculated
using a hydrodynamic simulation the rise to the peak in the light
curve for viewing angles outside the initial jet aperture
($\theta_{\rm obs} > \theta_0$) is much more gradual compared to a
semi-analytic model where the jet remains uniform with sharp edges
and no lateral expansion.\footnote{If lateral expansion is included in
such a semi-analytic model the rise to the peak in the light curve for
$\theta_{\rm obs} > \theta_0$ becomes even steeper, since the beaming
cone of the jet approaches and eventually engulfs the line of sight
faster.} This is because of the mildly relativistic material at the
sides of the jet whose emission is not strongly beamed away from such
off-axis lines of sight ($\theta_{\rm obs} > \theta_0$) at early
times, and therefore dominates the observed flux early on.

For a Gaussian jet, if both $\epsilon(\theta)$ and $\Gamma_0(\theta)$
have a Gaussian profile (corresponding to a constant rest mass per
solid angle in the outflow), then the afterglow light curves are
rather similar to those for a uniform jet \citep{KG03}. If, on the
other hand, $\epsilon(\theta)$ is Gaussian while\footnote{This
corresponds to a Gaussian angular distribution of the rest mass per
solid angle, i.e., very little mass near the outer edge of the jet,
which is the opposite of what might be expected due to mixing near the
walls of the funnel in the massive star progenitor.} $\Gamma_0(\theta)
= {\rm const}$, then the light curves for off-axis viewing angles
(i.e., outside the core of the jet) have a much higher flux at early
times, compared to a Gaussian $\Gamma_0(\theta)$ or a uniform jet (see
the bottom two panels of Figure \ref{fig:off-axis_LC}), due to a
dominant contribution from the emitting material along the line of
sight which has an early deceleration time in this case
\citep{GR-RP05}. Such a jet structure was considered as a
quasi-universal jet model \citep{Zhang04}.

\begin{figure}[!t]
\includegraphics[width=1.0\columnwidth]{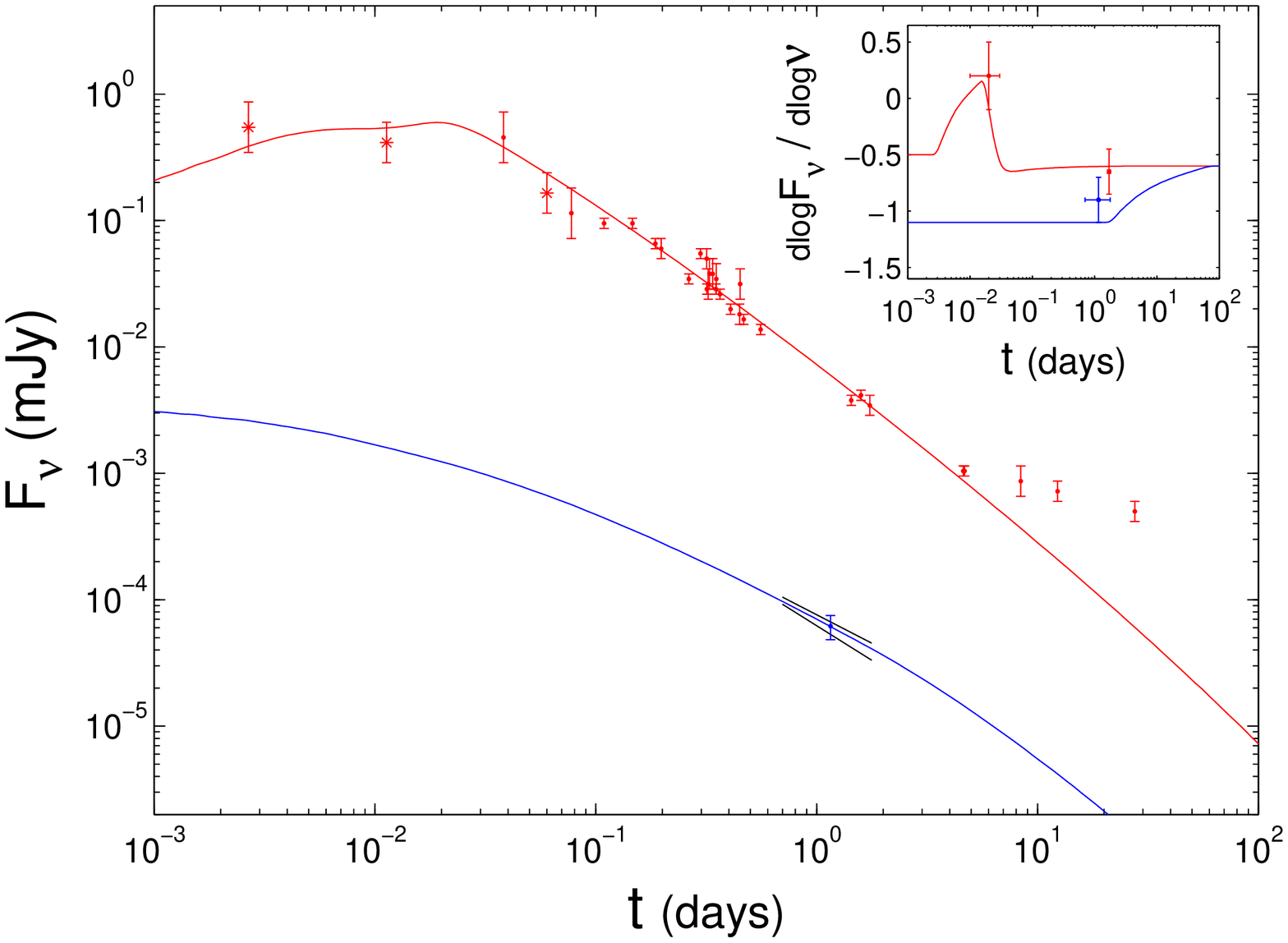}
\\
\vspace{-0.2cm}
\\
\includegraphics[width=1.0\columnwidth]{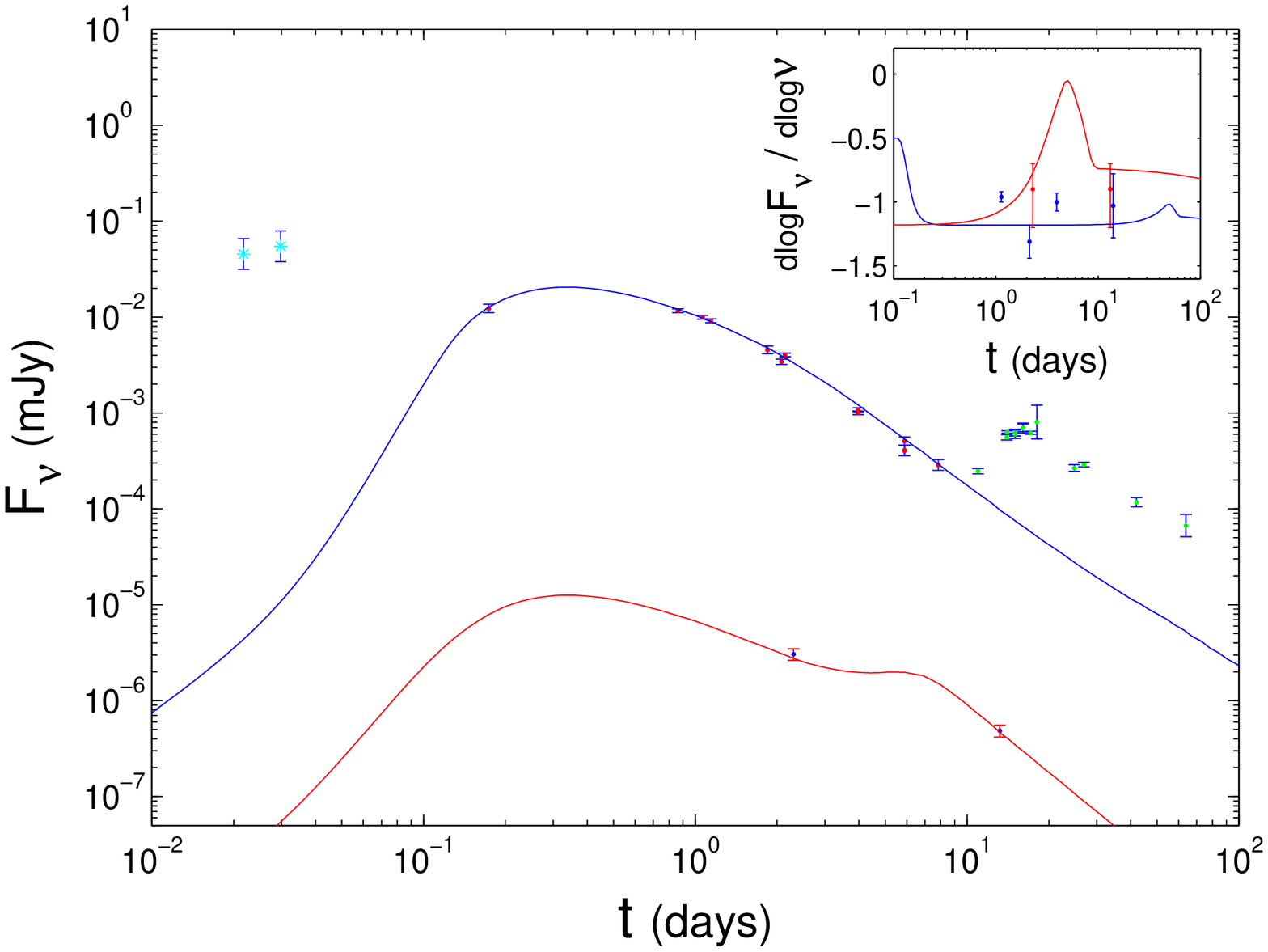}
\vspace{-0.8cm}
\\
\caption{Tentative fits to the afterglow light curves of X-ray rich
 GRB~041006 ({\it upper panel}) and X-ray flash XRF~030723 ({\it lower
 panel}) for a uniform sharp edged jet with a constant half-opening
 angle $\theta_0$, viewed from an angle $\theta_{\rm obs}$ slightly
 outside of its edge, with $(\theta_{\rm obs} - \theta_0) \sim
 0.15\theta_0$ for XRGRB~041006 and $(\theta_{\rm obs} - \theta_0)
 \sim \theta_0$ for XRF~030723 \citep[for details see][]{GR-RP05}. In
 both panels the upper curve is optical (R-band) and the lower curve
 is X-rays ($0.5-6\;$keV for XRGRB~041006 and $0.5-8\;$keV for
 XRF~030723), while the inset shows the optical and X-ray spectral
 slopes.}
\label{fig:XRF}
\end{figure}

\begin{figure}[!t]
\vspace{0.035cm}
\includegraphics[width=1.00\columnwidth]{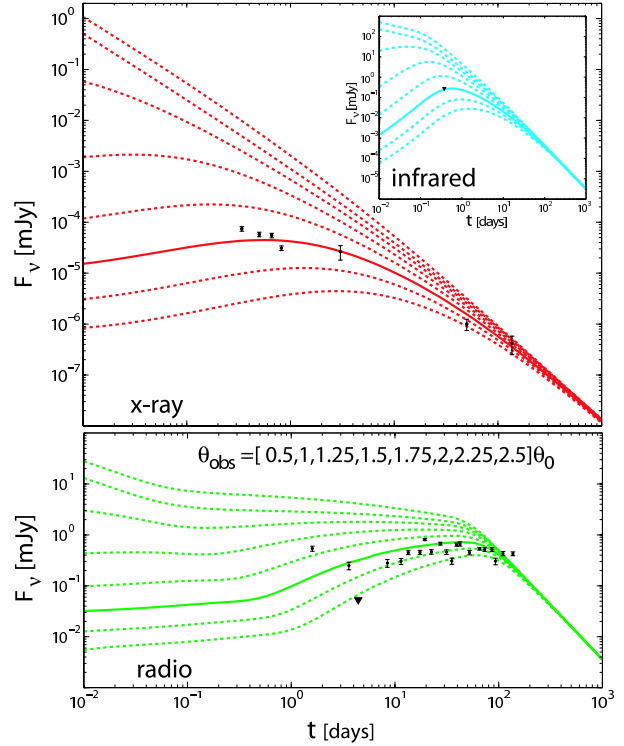}
\caption{A tentative fit to the afterglow light curve of GRB~031203
  with a uniform sharp edged jet of half-opening angle $\theta_0 =
  5^\circ$ for different viewing angles $\theta_{\rm obs}$ from its
  symmetry axis \citep[from][]{R-R05}.}
\label{fig:GRB031203}
\end{figure}

It has been suggested that a uniform jet with sharp edges viewed from
slightly outside of its edge ($\theta_0 < \theta_{\rm obs} \lesssim
2\theta_0$) would result in an X-ray flash (XRF) or X-ray rich GRB
(XRGRB) \citep{YIN02,YIN03,YIN04}, because of the smaller blue-shift
of the prompt emission compared to viewing angles within the jet
($\theta_{\rm obs} < \theta_0$). This has also been found to nicely
explain the flat early part of the light curve of XRGRB 041006 with
$(\theta_{\rm obs} - \theta_0) \sim 0.15\theta_0$ and XRF 030723 with
$(\theta_{\rm obs} - \theta_0) \sim \theta_0$ \citep[][see
Figure~\ref{fig:XRF}]{GR-RP05}, as naturally expected in this model,
since a larger angular displacement outside the edge of the jet should
result in a softer prompt emission.  There is a good case for a
viewing angle slightly outside the edge of a uniform sharp edged jet
also in the case of GRB~031203 \citep[][see
Figure~\ref{fig:GRB031203}]{R-R05}.

This expectation arises, however, assuming that the regions of
prominent gamma-ray emission and afterglow emission coincide. If this
assumption is relaxed \citep{EG06}, then an initially flat light curve
may appear in hard and bright GRBs, for lines of sight along which
there is bright gamma-ray emission but hardly any afterglow emission.
In this picture viewing angles outside the region within the jet with
bright afterglow emission can naturally account for the early flat
part of the X-ray afterglows detected by {\it Swift}
\citep{Nousek06}. These early flat decay stages cannot be attributed
to viewing angle effects in the USJ model, where there is always
significant afterglow emission along the line of sight.\footnote{With
the possible exception of very large viewing angles, $\theta_{\rm obs}
> \theta_{\rm max} \gtrsim 0.5\;$rad, if the jet has a sharp outer
edge at an angle $\theta_{\rm max}$, however in this case the event
would be very dim, and this cannot account for most of the {\it Swift}
GRBs with an early flat decay phase.} In scenarios where a break in
the light curve is caused by the jet structure and/or dynamics, the
break is expected to be largely achromatic.

\subsection{Orphan Afterglows}
\label{subsec:orphans}

For a jet with reasonably sharp edges, the prompt GRB emission becomes
very dim at viewing angles significantly outside the edge of the jet
($\Gamma\Delta\theta_{\rm obs} \gg 1$), and it can therefore be
detected only from within or slightly outside the initial jet
aperture, $\Gamma_0(\theta_{\rm obs}-\theta_0) \lesssim {\rm a\ few}$.
From larger viewing angles the prompt emission would not be detected.
During the afterglow, however, the jet decelerates and its beaming
cone widens with time, so that the afterglow emission at late times
can become detectable out to much larger viewing angles. Such events,
with no detected prompt GRB emission but with detected afterglow
emission in lower frequencies at later times, are called orphan
afterglows. While no such orphan afterglow has been detected to date,
their potential for constraining the degree of collimation of the jet
has been realized early on \citep{Rhoads97}.

In addition to ``off-axis'' orphan afterglows, where the prompt
emission is not detected due to a viewing angle outside of the jet,
there can also be ``on-axis'' orphan afterglows, where there is
significant afterglow emission along the line of sight from relatively
early times, but for some reason the prompt gamma-ray emission along
the line of sight is very weak \citep{HDL02,NP03}. This could occur,
for example, if the initial Lorentz factor along the line of sight is
not large enough to avoid excessive pair production (the
``compactness'' problem), $\Gamma_0 \lesssim 100$, while it is still
sufficiently large in order for the deceleration time $t_{\rm dec}$ to
be early enough to enable the detection of the afterglow emission
(typically $t_{\rm dec} \lesssim 1\;$day for $\Gamma_0 \gtrsim 10$).

Many works have analyzed the expected detection rate of orphan
afterglows, as well as the constraints on the degree of collimation of
GRB jets from existing surveys, in the X-ray \citep{WL99,NP03},
optical \citep{DGP02,TP02,NPG02,Rhoads03,RGS06}, and radio
\citep{PL98,Levinson02,GalYam06}. The constraints derived in this way
are still not very severe, but are nevertheless becoming increasingly
more interesting. Future surveys for orphan afterglows in the X-ray,
optical and radio can help constrain the degree of collimation of GRB
jets, as well as the jet structure. The detection of orphan GRB
afterglows would provide an important independent line of evidence in
favor of jets in GRBs.

\subsection{Some Implications of recent {\it Swift} Observations}
\label{subsec:Swift}

The ability of the {\it Swift} satellite to rapidly and autonomously
slew toward GRBs and observe them in X-rays, UV, and optical, has
dramatically improved our knowledge of the early afterglow
emission. In particular it provided excellent coverage of the early
X-ray afterglow, and enable rapid followup observations in the optical
and NIR by ground based robotic telescopes. In the context of jets,
there are two main new observations which are most relevant.

The first is the lack of a clear jet break in the afterglow light
curve of most {\it Swift} GRBs. Even when a break in the light curve
does show up, it is often chromatic \citep{Panaitescu06}, i.e. seen in
the X-rays but not in the optical, in stark contrast with the largely
achromatic nature that is expected for a jet break. The lack of a
clear jet break in many {\it Swift} GRBs can at least in part be
attributed to the larger sensitivity of {\it Swift} compared to
previous missions, that causes it to detect dimmer GRBs on average,
which in turn correspond to wider jets with a later jet break time and
lower flux at that time, making it harder to observe a clear jet
break.  It is not yet clear whether this explanation can fully account
for paucity of clear jet breaks in {\it Swift} GRBs. Furthermore, the
chromatic breaks that are seen in the afterglow light curves of some
{\it Swift} GRBs definitely require a novel explanation.

The second new {\it Swift} observation that bears relevance for the
efficiency of the prompt gamma-ray emission, $\epsilon_\gamma$, and
for the kinetic energy, $E_k$, of the jet during the late phases of
the afterglow for different jet structures \citep[following][]{GKP06},
is the flat decay phase in the early X-ray afterglow of many {\it
Swift} GRBs. Pre-{\it Swift} studies \citep{PK02,Yost03,L-RZ04} found
that the isotropic equivalent kinetic energy in the the afterglow
shock at late times (typically evaluated at $t = 10\;$hr), $E_{\rm
k,iso}(10\;{\rm hr})$, is comparable to the isotropic equivalent
energy output in gamma rays, $E_{\rm \gamma,iso}$, i.e. that typically
$\kappa \equiv E_{\rm \gamma,iso}/E_{\rm k,iso}(10\;{\rm hr}) \sim
1$. The gamma-ray efficiency is given by $\epsilon_\gamma = E_{\rm
\gamma,iso}/(E_{\rm \gamma,iso}+E_{\rm k,iso,0})$, where $E_{\rm
k,iso,0}$ is the initial value of $E_{\rm k,iso}$ corresponding to
material with a sufficiently large initial Lorentz factor ($\Gamma_0
\gtrsim 10^2$) that could have contributed to the prompt gamma-ray
emission. This implies a simple relation,
$\epsilon_\gamma/(1-\epsilon_\gamma) = \kappa f$, where $f \equiv
E_{\rm k,iso}(10\;{\rm hr})/E_{\rm k,iso,0}$ can be estimated from the
early afterglow light curve.

If the flat decay phase in the early X-ray afterglow observed by {\it
Swift} is interpreted as energy injection
\citep{Nousek06,Zhang06,Panaitescu06,GK06} this typically implies $f
\gtrsim 10$ and therefore $\epsilon_\gamma \gtrsim 0.9$. This is a
very high efficiency for any reasonable model for the prompt emission,
and in particular for the popular internal shocks model. If the early
flat decay phase is not due to energy injection, but is instead due to
an increase with time in the afterglow efficiency, then $f \sim 1$ and
typically $\epsilon_\gamma \sim 0.5$.  This is a more reasonable
efficiency, but still rather high for internal shocks. Such an
increase in the afterglow efficiency can occur, e.g., if one or more
of the following shock micro-physics parameters increases with time:
the fraction of the internal energy in relativistic electrons,
$\epsilon_e$, or in magnetic fields, $\epsilon_B$, or the fraction
$\xi_e$ of the electrons that are accelerated to a relativistic
power-law distribution of energies. If, in addition, $E_{\rm
k,iso}(10\;{\rm hr})$ had been underestimated, e.g. due to the
assumption that $\xi_e = 1$, then\footnote{\citet{EW05} have pointed
out a degeneracy where the same afterglow observations are obtained
under the substitution $(E,n) \to (E,n)/\xi_e$ and
$(\epsilon_e,\epsilon_B) \to \xi_e(\epsilon_e,\epsilon_B)$ for a value
of $\xi_e$ in the range $m_e/m_p \leq \xi_e \leq 1$, instead of the
usual assumption of $\xi_e = 1$.} $\kappa \sim \xi_e$ and $\xi_e \sim
0.1$ would lead to $\kappa \sim 0.1$ and $\epsilon_\gamma \sim 0.1$.

The internal shocks model can reasonably accommodate gamma-ray
efficiencies of $\epsilon_\gamma \lesssim 0.1$, which in turn imply
$\kappa \lesssim 0.1$. Since the true (corrected for beaming) energy
output in gamma rays, $E_\gamma = f_b E_{\rm\gamma,iso}$ where $f_b =
(1-\cos\theta_0) \approx \theta_0^2/2$, is clustered around
$10^{51}\;$erg \citep{Frail01,BFK03}, this implies $E_k(10\;{\rm hr})
= f_b E_{\rm k,iso}(10\;{\rm hr}) = E_\gamma/\kappa \gtrsim
10^{52}\;$erg for a uniform jet. For a structured jet with equal
energy per decade in $\theta$ ($\epsilon \propto \theta^{-2}$) in the
wings, the true energy in the jet is larger by a factor of
$1+2\ln(\theta_{\rm max}/\theta_c) \sim 10$, which implies
$E_k(10\;{\rm hr}) \gtrsim 10^{53}\;$erg in order to achieve
$\epsilon_\gamma \lesssim 0.1$. Such energies are comparable (for the
UJ model) or even higher (for the USJ model) than the estimated
kinetic energy of the Type Ic supernova (or hypernova) that
accompanies the GRB. This is very interesting for the total energy
budget of the explosion.

\section{Summary and Conclusions}
\label{sec:conc}

Our current understanding of GRB jets has been reviewed with special
emphasis on the jet dynamics (\S \ref{sec:dyn}) and structure (\S
\ref{sec:str}). The main conclusions are as follows. Semi-analytic
models predict a very rapid sideways expansion of the jet once its
Lorentz factor drops below the inverse of its initial half-opening
angle (\S \ref{subsec:semi-analytic}). Numerical studies, however,
show a very modest degree of lateral expansion as long as the jet is
relativistic. Such numerical studies include both an intermediate
approach where the hydrodynamic variables are integrated over the
radial profile of the jet, which significantly simplifies the
hydrodynamic equations (\S \ref{subsec:2D1D}), and full hydrodynamic
simulations (\S \ref{subsec:sim}).

The full hydrodynamic simulations are the most reliable of these
methods. The fact that the result of the intermediate method
(described in \S \ref{subsec:2D1D}) for an initially Gaussian jet
agree rather well with hydrodynamic simulations of an initially
uniform jet with sharp edges, lends credence to its results, which
show that also for a ``structured'' jet the lateral expansion is very
small (even smaller than for an initially uniform or Gaussian jet,
because of the smaller gradients in the lateral direction) and the
distribution of energy per solid angle remains very close to its
initial form as long as the jet is relativistic.

The afterglow image is expected to be rather uniform at low
frequencies (radio) and more limb brightened at higher frequencies
(optical, UV or X-rays). Its morphology and the evolution of its size
can help probe the jet dynamics and structure, as well as the external
density profile (\S~\ref{subsec:image}).

The observed jet break in the light curve in a uniform jet occurs
predominantly due to the lack of contribution to the observed flux
from outside the edges of the jet, once they become visible, and the
very modest lateral expansion does not play an important role (\S
\ref{subsec:break}).

The most popular models for the jet structure are the uniform jet (UJ)
model, where the jet is uniform within some finite half-opening angle
and has sharp edges, and the universal structured jet (USJ) model,
where the jet has a narrow core and wings where the energy per solid
angle drops as a power law (usually assumed to be an inverse square)
with the angle from the jet symmetry axis. There are also other jet
structures that have been discussed in the literature, which include a
Gaussian jet, a two component jet with a narrow uniform core of
initial Lorentz factor $\Gamma_0 \gtrsim 10^2$ surrounded by a wider
uniform component with $\Gamma_0 \sim 10-30$, and jets with a cross
section in the shape of a ring (or ``hollow cone'') or a fan (see
Figure \ref{fig:jet_structures}). 

There are various approaches for constraining the jet
structure. Statistical studies of the prompt emission are not very
conclusive yet, while the observed combined distribution of jet break
times and redshifts appears to disfavor the USJ model (\S
\ref{subsec:statistics}). The evolution of the linear polarization of
the afterglow provides interesting constraints on the jet structure
and the magnetic field configuration in the emitting region (\S
\ref{subsec:lin_pol}), but it is difficult to obtain tight constraints
on the jet structure without making strong assumptions about the
magnetic field configuration.

The shape of the afterglow light curves is an important and relatively
robust diagnostic tool for constraining the jet structure (\S
\ref{subsec:afterglow_LC}). It can practically rule out a jet with a
cross section of a narrow ring or a fan, and it constrains the
properties of a two component jet. The light curves for viewing angles
slightly outside the (reasonably sharp) edge of a jet would initially
rise or at least be very flat before they join the decaying light
curve for lines of sight within the (initial) jet aperture (\S
\ref{subsec:off-axis}). This can naturally account for such a flat
decay phase observed in the best monitored pre-{\it Swift} X-ray flash
(030723) and X-ray rich GRB (041006). It can also explain the early
X-ray afterglow light curves of many {\it Swift} GRBs if the regions
of prominent gamma-ray and afterglow emission do not coincide
\citep{EG06}. However, this cannot be attributed to viewing angle
effects in the USJ model, while in the Gaussian jet model it suggests
that the initial Lorentz factor significantly drops outside of the jet
core.

Our understanding of the structure and dynamics of GRB jets may
improve thanks to future observations. These include a dense
monitoring of the afterglow emission starting at early times and over
a wide range of frequencies (radio, mm, NIR, optical, UV, X-ray),
polarization measurements with good temporal coverage both at early
times (much earlier than the jet break time) and around (i.e. at least
a factor of a few before and after) the jet break time, and measuring
the evolution of the afterglow image size. Surveys for orphan
afterglows (\S \ref{subsec:orphans}) are already beginning to provide
interesting constraints on the collimation of GRB jets, and future
surveys may also constrain the jet structure. The detection of orphan
GRB afterglows may also provide an important independent line of
evidence for jets in GRBs. Recent {\it Swift} observations (\S
\ref{subsec:Swift}) show a paucity of clear achromatic jet breaks in
the afterglow light curves as well as chromatic breaks which challenge
existing models and call for new explanations.

\acknowledgements

I am grateful to Ehud Nakar, Enrico Ramirez-Ruiz, and Arieh K\"onigl
for useful comments on the manuscript. This research was supported by
the US Department of Energy under contract number DE-AC03-76SF00515.

\end{document}